\documentclass[11pt]{article}

\usepackage[]{inputenc}
\usepackage[T1]{fontenc}
\usepackage{hyperref}

\usepackage{color}

\usepackage{fullpage}

\usepackage{amsmath}
\usepackage{amssymb}
\usepackage[toc,page]{appendix}

\usepackage{amsthm}
\usepackage{float}
\usepackage{wrapfig}
\usepackage{caption}
\usepackage{graphicx}
\usepackage{pgfplots}
\usepackage{subcaption}
\usepackage{tikz,tikz-3dplot}
\tdplotsetmaincoords{80}{45}
\tdplotsetrotatedcoords{-90}{180}{-90}

\tikzset{surface1/.style={draw=blue!70!black, fill=blue!40!white, fill opacity=.6}}
\tikzset{surface2/.style={draw=red!70!black, fill=red!40!white, fill opacity=.6}}
\tikzset{surface3/.style={draw=green!70!black, fill=green!40!white, fill opacity=.6}}

\newcommand{\coneback}[4][]{
  \draw[canvas is xy plane at z=#2, #1] (45-#4:#3) arc (45-#4:225+#4:#3) -- (O) --cycle;
  }
\newcommand{\conefront}[4][]{
  \draw[canvas is xy plane at z=#2, #1] (45-#4:#3) arc (45-#4:-135+#4:#3) -- (O) --cycle;
  }

\usetikzlibrary{intersections,fadings,decorations.pathreplacing}

  \pgfplotsset{
        compat=1.8}

\theoremstyle{definition}

\def \be {\begin{equation}}
\def \ee {\end{equation}}
\def \bea {\begin{eqnarray}}
\def \eea {\end{eqnarray}}

\title{Well-posed formulation of Einstein-Maxwell effective field theory}

\def\be{\begin{equation}}
\def\ee{\end{equation}}
\def\ba{\begin{eqnarray}}
\def\ea{\end{eqnarray}}

\author{Iain Davies and Harvey S. Reall\\ \\ {\small Department of Applied Mathematics and Theoretical Physics, University of Cambridge}\\ {\small Wilberforce Road, Cambridge CB3 0WA, United Kingdom} \\ id318@cam.ac.uk, hsr1000@cam.ac.uk}

\begin{document}
\maketitle

\begin{abstract}
We consider the well-posedness of the initial value problem for Einstein-Maxwell theory modified by higher derivative effective field theory corrections. Field redefinitions can be used to bring the leading parity-symmetric 4-derivative corrections to a form which gives second order equations of motion. We show that a recently introduced ``modified harmonic'' gauge condition can be used to obtain a formulation of these theories which admits a well-posed initial value problem when the higher derivative corrections to the equations of motion are small.
\end{abstract}

\section{Introduction}

The only known fundamental fields for which the classical approximation is useful are the gravitational and electromagnetic fields. To an excellent approximation these are described by conventional Einstein-Maxwell theory but it is known that this theory will be modified by higher order effective field theory (EFT) corrections. In this paper we will consider the initial value problem for Einstein-Maxwell theory with the leading order EFT corrections. 

In EFT we write the action as an expansion involving terms with increasing numbers of derivatives:\footnote{
Dimensional analysis suggests that this should be viewed as a double expansion ordered by increasing number of $F_{\mu\nu}$ factors, and increasing number of derivatives.}
\be
 S = \int d^4 x \sqrt{-g} \left[\frac{1}{16\pi G}\left(  -2\Lambda + R \right) - \frac{1}{4} F_{\mu\nu} F^{\mu\nu} + L_4 + L_6 + \ldots \right]
\ee
where $F=dA$ with $A_\mu$ the vector potential and  the scalar $L_n$ is a polynomial in derivatives of the fields, where each term contains a total of $n$ derivatives of the fields $(g_{\mu\nu},A_\rho)$. For example $L_4$ contains terms such as $R^2$, $R_{\mu\nu} R^{\mu\nu}$ and $(F_{\mu\nu}F^{\mu\nu})^2$.

If we truncate the above theory by discarding terms with $6$ or more derivatives then the resulting 4-derivative theory describes the leading EFT corrections to conventional Einstein-Maxwell theory. However, $L_4$ gives terms in the equations of motion containing third or fourth derivatives of the fields $(g_{\mu\nu},A_\rho)$. This is problematic for two reasons. First, the mathematical properties of the equations of motion are very sensitive to the terms with the most derivatives. If these terms do not have a nice algebraic structure then the initial value problem will not be well-posed. Second, even if these equations admit a well-posed formulation, as is the case for 4-derivative corrections to pure gravity in 4d \cite{Noakes:1983xd}, the higher order nature of the equations of motion means that additional initial data are required, which means that the equations describe spurious (massive) degrees in addition to the two fields present in the EFT. 

One way around these problems is to treat the higher derivative terms perturbatively, i.e., construct solutions as expansions in the coefficients of the higher derivative terms. However, there are situations where such expansions exhibit secular growth, leading perturbation theory to break down in a situation when EFT should remain valid \cite{Flanagan:1996gw}. If a formulation of the equations could be found that admits a well-posed initial value problem then we would not be restricted to constructing solutions perturbatively. 

To make progress, we exploit the fact that in EFT, the higher derivative terms in the Lagrangian are not unique, but can be adjusted order by order in using field redefinitions. This enables one to freely adjust the coefficients of terms in the Lagrangian that are proportional to equations of motion. For example, in the case of pure gravity, the coefficients of the terms $R^2$ and $R_{\mu\nu}R^{\mu\nu}$ can be adjusted by a field redefinition of the form $g_{\mu\nu} \rightarrow g_{\mu\nu} +  d_1 R_{\mu\nu} +  d_2 R g_{\mu\nu} + \ldots$ with suitable choices of $d_1,d_2$. This can be used to make $L_4$ proportional to the Euler density (of the Gauss-Bonnet invariant). In 4d this term is topological, i.e., it does not affect the equations of motion, and so this shows that one can eliminate 4-derivative corrections in 4d pure gravity.

Similarly, in Einstein-Maxwell theory, field redefinitions can be used to bring $L_4$ to the form (neglecting topological terms)
\be
\label{L4}
 L_4 = c_1 X^2 + c_2 Y^2 + c_3 R_{\mu\nu\rho\sigma} \tilde{F}^{\mu\nu} \tilde{F}^{\rho\sigma} + c_4 XY + c_5 R_{\mu\nu\rho\sigma} F^{\mu\nu} \tilde{F}^{\rho\sigma}
\ee
where
\be
 \tilde{F}_{\mu\nu} = \frac{1}{2} \epsilon_{\mu\nu\rho\sigma} F^{\rho\sigma}
\ee
and
\be
X = F_{\mu\nu} F^{\mu\nu} \qquad \qquad Y = F_{\mu\nu} \tilde{F}^{\mu\nu}
\ee
The terms with coefficients $c_1$, $c_2$ and $c_3$ are symmetric under space-time orientation reversal (i.e. under parity or time-reversal) whereas the terms with coefficients $c_4$ and $c_5$ break this symmetry. In the parity-symmetric case, the above form of the Lagrangian can be determined from results in \cite{Deser:1974cz}. Ref. \cite{Jones:2019nev} discusses the parity violating terms (for a more general class of theories).

It is well-known that $c_1,c_2,c_3$ receive contributions from QED effects. In flat spacetime, at energies well below the electron mass $m$, QED predicts corrections to Maxwell theory described by the Euler-Heisenberg EFT which has specific values for $c_1$ and $c_2$ proportional to $\alpha^2/m^4$ where $\alpha$ is the fine-structure constant. In curved spacetime, the term with coefficient $c_3$ also arises from ``integrating out" the electron in this way, with $c_3 \propto \alpha/m^2$ \cite{Drummond:1979pp}. 

The nice thing about using field redefinitions to write $L_4$ as above is that all of the terms except for the last one give rise to second order equations of motion (for the $c_3$ term this follows from \cite{Horndeski:1980}). In particular, if we restrict to a theory with $c_5=0$ (e.g. a parity symmetric theory) then the equations of motion are second order and we can hope that the theory admits a well-posed initial value problem. 

If we ignore gravity and just consider the 4-derivative corrections to Maxwell theory (the terms quadratic in $X,Y$) then it can be shown that the equations of motion can be written as a first order symmetric hyperbolic system for $F_{\mu\nu}$, which ensures a well-posed initial value problem \cite{Abalos:2015gha}. This result holds only when the 4-derivative corrections are small, i.e., within the regime of validity of EFT. 

With dynamical gravity, well-posedness is much more complicated. There are no gauge-invariant observables for gravity so any formulation of the equations requires a choice of gauge. By ``formulation" we mean a choice of gauge plus a way of  gauge-fixing the equations. Even for the 2-derivative vacuum Einstein equation it is well-known that many formulations do not give a well-posed initial value problem (the same is true for the Maxwell equations viewed as equations for $A_\mu$). For example, the ADM formulation of the Einstein equation is only weakly hyperbolic \cite{Kidder:2001tz}, which is not enough to ensure a well-posed initial value problem. Choquet-Bruhat \cite{Bruhat:1952} was the first to show that a well-posed formulation existed by proving the harmonic gauge formulation met this criteria. There are also modifications of the ADM formulation, such as the BSSN formulation \cite{Baumgarte:1998te,Shibata:1995we}, that are strongly hyperbolic \cite{Sarbach:2002bt,Nagy:2004td} and therefore do admit a well-posed initial value problem. 

Even if one considers a formulation of the equations of motion that gives a well-posed initial value problem for 2-derivative Einstein-Maxwell theory, there is no guarantee that well-posedness will persist when one deforms the theory to include 4-derivative corrections, no matter how small. This has been seen in recent work on the EFT of gravity coupled to a scalar field. 
In this EFT, one considers gravity minimally coupled to the scalar field and then extends this 2-derivative theory by including 4-derivative corrections. Field redefinitions can be used to write (parity symmetric) 4-derivative terms in a form that gives second order equations of motion. The simplest strongly hyperbolic formulation of the 2-derivative theory is based on harmonic gauge. However, if one includes 4-derivative corrections then this formulation is only weakly hyperbolic, even for arbitrarily small 4-derivative terms, so the initial value problem is not well-posed \cite{Papallo:2017qvl,Papallo:2017ddx}.\footnote{This EFT is a Horndeski theory, i.e., a diffeomorphism invariant scalar-tensor theory with second order equations of motion. Strongly hyperbolic BSSN-like formulations have been found for a certain subset of Horndeski theories \cite{Kovacs:2019jqj} but this subset does not include the EFT we are discussing.}

Fortunately it has been shown that there exists a deformation of the harmonic gauge formulation that {\it does} give strongly hyperbolic equations when the 4-derivative terms are small \cite{Kovacs:2020pns,Kovacs:2020ywu}. This smallness requirement is not a concern because it is also required for the validity of EFT (if higher derivative terms are not small then a ``UV" description of the physics would be necessary). This ``modified harmonic gauge" formulation gives a well-posed initial value problem for the gravity-scalar EFT in 4d, as well as for the EFT of pure gravity in higher dimensions (where the Euler density is not topological). 

In this paper we will consider the theory \eqref{L4} with $c_5=0$, which describes Einstein-Maxwell theory with the leading (parity-symmetric) higher derivative EFT corrections. The similarity with the Einstein-scalar case strongly suggests that a conventional harmonic/Lorenz gauge formulation of this theory will be only weakly hyperbolic even when the $4$-derivative terms are small. However, we can adapt the modified harmonic gauge formulation to this EFT. We will show that the resulting equations are strongly hyperbolic when the 4-derivative terms are small, and therefore this formulation admits a well-posed initial value problem when it is within the regime of validity of EFT. Our results apply also to the larger class of ($c_5=0$) theories obtained by replacing the terms quadratic in $X,Y$ with an arbitrary smooth function $f(X,Y)$ satisfying $f(0,0)=f_X(0,0)=f_Y(0,0)=0$. This includes, for example, the Born-Infeld Lagrangian for nonlinear electrodynamics. 

This paper is organised as follows. In Section \ref{Strong Hyperbolicity} we review briefly the notion of strong hyperbolicity. In Section \ref{Setup of IVP} we
describe the modified harmonic gauge formulation and determine the principal symbol of the equations of motion. In Section \ref{Proof} we prove that these equations are strongly hyperbolic following arguments very close to those of \cite{Kovacs:2020ywu}. In Section \ref{Conclusion} we make a few concluding remarks. 

\section{Strong Hyperbolicity} \label{Strong Hyperbolicity}

A full overview of strong hyperbolicity and how it relates to well-posedness is given in \cite{Kovacs:2020ywu}. A summary is provided here for context.

Consider a second order PDE for a set of fields $u_I$, $I=1, \ldots, N$. In our case we will have $u_I =(g_{\mu\nu},A_\rho)$, and so $N=10 + 4 = 14$ (10 for the independent components of a 4 $\times$ 4 symmetric matrix and 4 for the components of a 4-vector). Assume that initial data is prescribed on some surface $\Sigma$. We pick a co-ordinate system $(x^0, x^i)$ with $\Sigma$ the surface $x^0=0$. The equations that we will consider are not quasilinear (linear in second derivatives) however they are linear in $\partial_0\partial_0 u$, i.e. the equations can be written in the form
    \begin{equation} \label{invert1}
        A^{IJ}(x,u,\partial_\mu u,\partial_0 \partial_i u, \partial_i \partial_j u)\partial_0^2 u_J = F^I(x,u,\partial_\mu u,\partial_0 \partial_i u, \partial_i \partial_j u)
    \end{equation}
Initial data must be chosen such that the slice $x_0 = 0$ is non-characteristic, meaning that $A^{IJ}$ is invertible there.
The {\it principal symbol} $\mathcal{P}(\xi)^{IJ}$ is a $N \times N$ matrix which determines the coefficients of the second-derivative terms when the equations of motion are linearized around a background (see Appendix for details). $\xi_\mu$ is an arbitrary covector. $\mathcal{P}(\xi)^{IJ}$ is quadratic in $\xi_\mu$ and so we can write it in the following form
\begin{equation} \label{eq:ABCdef}
    \mathcal{P}(\xi)^{IJ}=\xi_0^2 A^{IJ} +\xi_0 B(\xi_i)^{IJ}+ C(\xi_i)^{IJ}  
\end{equation}
Here $A^{IJ}$ is the same as in ($\ref{invert1}$) by the definition of $\mathcal{P}(\xi)^{IJ}$. The N by N matrices $A^{IJ}$, $B^{IJ}$ and $C^{IJ}$ have additional suppressed arguments $(x,u,\partial_\mu u,\partial_0 \partial_i u, \partial_i \partial_j u)$ but not $\partial_0\partial_0 u$ by the linearity condition.

We now define the matrix
\begin{equation} \label{Mmatrix}
    M(\xi_i) = \begin{pmatrix}
        0 & I\\
        -A^{-1}C & -A^{-1}B
    \end{pmatrix}
\end{equation}
Let $\xi_i$ be unit with respect to some smooth Riemannian metric $G^{ij}$ on surfaces of constant $x^0$. Then the system of equations ($\ref{invert1}$) is {\it strongly hyperbolic} if for any such $\xi_i$, there exists a positive definite matrix $K(\xi_i)$ that depends smoothly on $\xi_i$ and its other suppressed arguments $(x,u,\partial_\mu u,\partial_0 \partial_i u, \partial_i \partial_j u)$ such that
\begin{equation} \label{eq:symmet}
    KM = M^\dagger K
\end{equation}
and there exists a positive constant $\lambda$ such that
\begin{equation}
    \lambda^{-1}I<K<\lambda I
\end{equation}
$K(\xi)$ is called the "symmetrizer".

There is a theorem proved in Chapter 5 of \cite{Taylor:1991} that if a first order system can be written in the form
\begin{equation} \label{firstorder}
    \partial_0 u_J = B_J (x, u, \partial_i u),
\end{equation}
then well-posedness of its Cauchy problem follows from the system being strongly hyperbolic (for an appropriate definition of strong hyperbolicity for first order systems), assuming appropriate regularity of initial data. Now, \cite{Kovacs:2020ywu} explains a construction for reducing second order systems of the form (\ref{invert1}) to a first order system of the form (\ref{firstorder}) in such a way that strong hyperbolicity of the second order system implies strong hyperbolicity of the first order system, provided $M$ is invertible. Hence, we can prove the Cauchy problem is well-posed by proving strong hyperbolicity of our second order system. 

The condition that $M$ be invertible is necessary for the reduction to a first order system used in \cite{Kovacs:2020ywu}. It will be shown to be invertible for our Einstein-Maxwell EFT under certain conditions, as discussed at the end of Section \ref{Standard E-M}.

\section{Setup of Initial Value Problem} \label{Setup of IVP}
\subsection{Equations of Motion}

We shall consider a theory with action
\begin{equation} \label{action}
    S = \frac{1}{16\pi G} \int_M d^4 x \sqrt{-g} \left[ -2\Lambda + R - \frac{1}{4} F_{\mu\nu} F^{\mu\nu} + f(X,Y) + c_3 R_{\mu\nu\rho\sigma} \tilde{F}^{\mu\nu} \tilde{F}^{\rho\sigma} \right]
\end{equation}
where $f(0,0)=f_X(0,0)=f_Y(0,0)=0$. This includes our ($c_5=0$) Einstein-Maxwell EFT, where we rescaled the Maxwell field to scale out an onverall factor of $16 \pi G$. The equations of motion result from varying $S$ with respect to $A_\mu$ and $g_{\mu\nu}$. We define variations
\begin{equation}
    E^{\mu} \equiv -\frac{16 \pi G}{\sqrt{-g}} \frac{\delta S}{\delta A_\mu}, \,\,\,\,\,\,
    E^{\mu\nu} \equiv -\frac{16 \pi G}{\sqrt{-g}} \frac{\delta S}{\delta g_{\mu\nu}}
\end{equation}
Explicitly these are given by
\begin{multline}
    E^{\mu\nu}= \Lambda g^{\mu\nu} + G^{\mu\nu} +\frac{1}{2} \left( \frac{1}{4} g^{\mu\nu}F^{\rho\sigma}F_{\rho\sigma}-F^{\mu\rho}F^{\nu}_{\,\,\,\rho} \right)\\
        - \frac{1}{2}g^{\mu\nu}\left(f-Y\partial_Y f\right) + 2F^{\mu\alpha} F^{\nu}_{\,\,\,\alpha}\partial_X f\\
    - \frac{1}{2}c_3\, g^{\nu\alpha}\,\delta^{\mu\lambda\rho\sigma}_{\alpha\beta\gamma\delta}\,\left(F_{\lambda\tau}F^{\beta\tau}R_{\rho\sigma}^{\,\,\,\,\,\,\gamma\delta}+\nabla^{\delta}F_{\lambda\rho}\nabla_{\sigma}F^{\beta\gamma}\right)
\end{multline}
\begin{multline}
    E^\mu = \nabla_{\nu}F^{\mu\nu}( 1 - 4\partial_X f)\\ - 2 \nabla_\nu F_{\alpha\beta} \left( 4F^{\mu\nu}F^{\alpha\beta}\partial_X^2 f + 2\left( F^{\mu\nu}F_{\gamma\delta}\epsilon^{\alpha\beta\gamma\delta} + \epsilon^{\mu\nu\rho\sigma}F_{\rho\sigma}F^{\alpha\beta} \right)\partial_X\partial_Y f + \epsilon^{\mu\nu\rho\sigma}\epsilon^{\alpha\beta\gamma\delta}F_{\rho\sigma}F_{\gamma\delta}\partial^2_Y f\right)\\ + c_3\,\delta^{\mu\nu\rho\sigma}_{\alpha\beta\gamma\delta}\,\nabla_{\nu}F^{\alpha\beta}R_{\rho\sigma}^{\,\,\,\,\,\,\gamma\delta}
\end{multline}
Let $n_\mu$ be a normal to the initial surface $\Sigma$. In our coordinate chart we have $n_\mu \propto \delta^0_\mu$. The anti-symmetries of the generalized Kronecker delta mean that $n_\mu E^\mu$ and $n_\mu E^{\mu\nu}$ don't contain any second $x^0$ derivatives, and hence these are constraint equations, as in conventional Einstein-Maxwell theory. Furthermore, these anti-symmetries also imply these equations are linear in the second derivative with respect to any one co-ordinate, and hence can be put in the form ($\ref{invert1}$). This will remain true when we include the gauge-fixing terms described below.

\subsection{Bianchi Identities}

The action $S$ is invariant under diffeomorphisms and electromagnetic gauge transformations. By considering an infinitesimal diffeomorphism $x^\mu \rightarrow \Tilde{x}^\mu(x) = x^\mu - \epsilon^\mu(x)$ we get the Bianchi identity
\begin{equation} \label{eq:6}
    \nabla_\nu E^{\mu\nu}-\frac{1}{2}F^\mu_{\,\,\,\nu} E^\nu + \frac{1}{2}A^\mu \nabla_\nu E^\nu = 0
\end{equation}
Furthermore, by considering a gauge transformation $A_\mu \rightarrow A_\mu + \nabla_\mu \chi$ we get 
\begin{equation} \label{eq:bim}
    \nabla_\mu E^\mu = 0
\end{equation}
Hence we can rewrite (\ref{eq:6}) as
\begin{equation} \label{eq:big}
    \nabla_\nu E^{\mu\nu}-\frac{1}{2}F^\mu_{\,\,\,\nu} E^\nu = 0
\end{equation}
These equations hold for any field configuration. They will allow us to prove the propagation of the modified harmonic gauge condition. They will also have consequences for the principal symbol discussed below.

\subsection{Modified Harmonic Gauge}

\begin{figure}[H]
\centering
\begin{subfigure}{0.5\linewidth}
\centering
\begin{tikzpicture}[tdplot_main_coords,scale=1.0]
  \coordinate (O) at (0,0,0);
  \coneback[surface3]{-3}{3.5}{-15}
  \coneback[surface2]{-3}{3}{-10}
  \coneback[surface1]{-3}{2.5}{-5}
  \conefront[surface3]{-3}{3.5}{-15}
  \conefront[surface2]{-3}{3}{-10}
  \conefront[surface1]{-3}{2.5}{-5}
  \coneback[surface3]{3}{3.5}{15}
  \coneback[surface2]{3}{3}{10}
  \coneback[surface1]{3}{2.5}{5}
  \conefront[surface3]{3}{3.5}{15}
  \conefront[surface2]{3}{3}{10}
  \conefront[surface1]{3}{2.5}{5}
  \node[above,color=blue!70!black] (1) at (0.5,2,3.3){$g^{\mu\nu}$};
  \node[above,color=red!70!black] (2) at (0.5,3,3.2){${\tilde g}^{\mu\nu}$};
  \node[above,color=green!70!black] (3) at (0.5,4,3.1){${\hat g}^{\mu\nu}$};
\end{tikzpicture}
\caption{}
\label{fig:cones1}
\end{subfigure}%
\begin{subfigure}{0.5\linewidth}
\centering
\begin{tikzpicture}[tdplot_main_coords,scale=1.0]
  \coordinate (O) at (0,0,0);
  \coneback[surface1]{-3}{2.5}{-15}
  \coneback[surface2]{-3}{2}{-10}
  \coneback[surface3]{-3}{1.5}{-5}
  \conefront[surface1]{-3}{2.5}{-15}
  \conefront[surface2]{-3}{2}{-10}
  \conefront[surface3]{-3}{1.5}{-5}
  \coneback[surface1]{3}{2.5}{15}
  \coneback[surface2]{3}{2}{10}
  \coneback[surface3]{3}{1.5}{5}
  \conefront[surface1]{3}{2.5}{15}
  \conefront[surface2]{3}{2}{10}
  \conefront[surface3]{3}{1.5}{5}
  \node[above,color=blue!70!black] (1) at (0.5,3,3.3){$g_{\mu\nu}$};
  \node[above,color=red!70!black] (2) at (0.5,1.5,3.4){$({\tilde g}^{-1})_{\mu\nu}$};
  \node[above,color=green!70!black] (3) at (0.5,-0.5,3.6){$({\hat g}^{-1})_{\mu\nu}$};
\end{tikzpicture}
\caption{}
\label{fig:cones2}
\end{subfigure}
 Reproduced from \cite{Kovacs:2020ywu} with permission of authors. \caption{(a) Cotangent space at a point, showing the
null cones of $g^{\mu\nu}$, $\tilde{g}^{\mu\nu}$ and $\hat{g}^{\mu\nu}$. 
(b) Tangent space at a point, showing the null cones of $g_{\mu\nu}$, $(\tilde{g}^{-1})_{\mu\nu}$ and $(\hat{g}^{-1})_{\mu\nu}$.}
\label{fig:cones}
\end{figure}
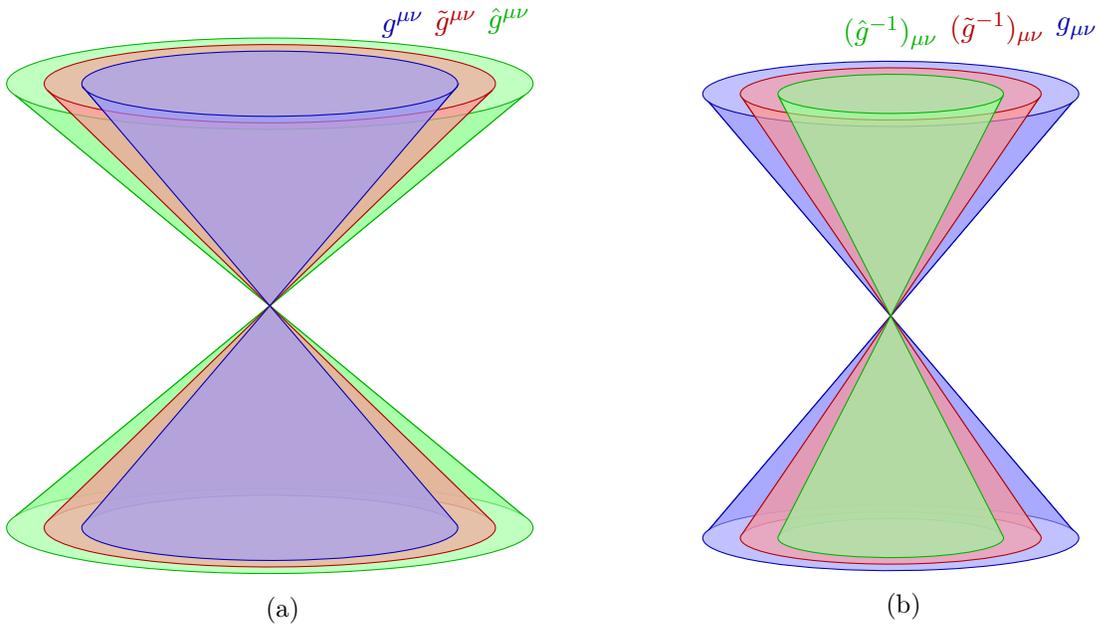

The "modified harmonic" gauge introduced in \cite{Kovacs:2020ywu} requires the introduction of two auxiliary metrics, $\tilde{g}^{\mu\nu}$ and $\hat{g}^{\mu\nu}$. These are completely unphysical, introduced only for purposes of fixing the gauge. Any index raising and lowering is done using the physical metric as usual. 

The only properties we require of the auxiliary metrics are that the (cotangent space) null cones of $\tilde{g}^{\mu\nu}$,  $\hat{g}^{\mu\nu}$ and $g^{\mu\nu}$ are nested as shown in Fig. \ref{fig:cones}(a), with the the null cone of $g^{\mu\nu}$ lying inside the null cone of $\tilde{g}^{\mu\nu}$, which lies inside the null cone of $\hat{g}^{\mu\nu}$.\footnote{
Ref. \cite{Kovacs:2020ywu} discusses alternative choices for the ordering of the nested cones.
} This nested structure ensures that any covector that is causal with respect to $g^{\mu\nu}$ is timelike with respect to $\tilde{g}^{\mu\nu}$ or $\hat{g}^{\mu\nu}$. This implies that if $\Sigma$ is spacelike with respect to $g^{\mu\nu}$, then it is also spacelike with respect to $\tilde{g}^{\mu\nu}$ and $\hat{g}^{\mu\nu}$. Ref. \cite{Kovacs:2020ywu} provides examples for how to construct such auxiliary metrics, such as $\tilde{g}^{\mu\nu} = g^{\mu\nu} - a n^\mu n^\nu$, $\hat{g}^{\mu\nu} = g^{\mu\nu} - b n^\mu n^\nu$ where $n^\mu$ is the unit normal to surfaces of constant $x^0$ and $a,b$ are functions chosen to take values in a certain range. In the tangent space, the null cones are also nested, with the ordering reversed, as shown in Fig. \ref{fig:cones}(b).  

Modified harmonic gauge is defined by\footnote{
It would be more accurate to refer to this as ``modified harmonic/Lorenz" gauge, since it modifies the harmonic gauge condition on the metric and the Lorenz gauge condition on the Maxwell field. However, we will stick with the shorter name.}
\be
 H^\mu = H = 0
\ee
where
\begin{align}
    H^\mu &\equiv \tilde{g}^{\rho\sigma}\nabla_\rho \nabla_\sigma x^\mu\\
    H &\equiv -\tilde{g}^{\nu\sigma}\nabla_\sigma A_\nu
\end{align}
and the gauge-fixed equations of motion are taken to be
\begin{align}
    E^{\mu\nu}_{mhg} &\equiv E^{\mu\nu}+\hat{P}_{\alpha}^{\,\,\,\beta\mu\nu}\partial_\beta H^\alpha=0 \label{eq:gfg}\\
    E^\mu_{mhg} &\equiv E^\mu +\hat{g}^{\mu\nu}\nabla_\nu H=0 \label{eq:gfm}
\end{align}
where
\begin{equation}
    \hat{P}_{\alpha}^{\,\,\,\beta\mu\nu} = \delta_\alpha^{(\mu}\hat{g}^{\nu)\beta}-\frac{1}{2}\delta_\alpha^\beta \hat{g}^{\mu\nu}
\end{equation}
Standard harmonic/Lorenz gauge would result from choosing $\hat{g}^{\mu\nu} = \tilde{g}^{\mu\nu}=g^{\mu\nu}$. However, the similarity with the Einstein-scalar case \cite{Papallo:2017qvl,Papallo:2017ddx} strongly suggests that this would result in equations that are only weakly hyperbolic. 

Note that the gauge fixing terms are linear in second derivatives of $g_{\mu\nu}$ and $A_\rho$, and hence equations (\ref{eq:gfg}) and (\ref{eq:gfm}) can still be put in the form ($\ref{invert1}$). For standard 2-derivative Einstein-Maxwell theory in modified harmonic gauge, the matrix $A^{IJ}$ is invertible on surfaces of constant $x^0$ provided the surfaces are spacelike (i.e. spacelike surfaces are non-characteristic). We will require that our initial data is chosen such that $A^{IJ}$ is invertible, i.e. the surface $x^0=0$ is non-characteristic. By continuity (of $\det A^{IJ}$), this will be the case for a spacelike initial surface if the initial data is sufficiently {\it weakly coupled}, i.e., the higher derivative terms are small compared to the $2$-derivative terms. By continuity, invertibility of $A^{IJ}$ will continue to hold in a neighbourhood of $x^0=0$. 
Hence, for weakly coupled initial data, equations (\ref{eq:gfg}) and (\ref{eq:gfm}) meet the prerequisites to consider whether they are strongly hyperbolic. The notion of weak coupling will be defined more precisely in section \ref{Weak}. 

\subsection{Propagation of Gauge Condition}

We must show that a solution to the modified harmonic gauge equations is also a solution to the original equations of motion provided the initial data satisfies the constraint equations and gauge conditions. 
We can impose $H^\mu=0$ on our initial data surface $\Sigma$, by choosing co-ordinates $x^\mu$ as argued in \cite{Kovacs:2020ywu}. Additionally $H=0$ can be imposed on $\Sigma$ by gauge transformation of $A_\mu$. The propagation of these conditions follows from much the same argument as in \cite{Kovacs:2020ywu}. Suppose we have a solution to equations (\ref{eq:gfg}) and (\ref{eq:gfm}) that satisfies the constraint equations $n_\mu E^\mu = 0$ and $n_\mu E^{\mu\nu} = 0$ on $\Sigma$. Then by taking the divergence of (\ref{eq:gfm}) and using the Bianchi identity (\ref{eq:bim}) we have 
\begin{equation} \label{eq:divbianchi}
    0 = \nabla_{\mu}E^{\mu}_{mhg} = \hat{g}^{\mu\nu}\nabla_\mu \nabla_\nu H + (\nabla_\mu \hat{g}^{\mu\nu})\nabla_\nu H 
\end{equation}
which is a linear wave equation for $H$ with principal symbol defined by $\hat{g}^{\mu\nu}$. We know that $\Sigma$ is spacelike with respect to $\hat{g}^{\mu\nu}$ and hence this equation has a well-posed initial value formulation. This means it has a unique solution in $\hat{D}(\Sigma)$ (the domain of dependence of $\Sigma$ with respect to $\hat{g}^{\mu\nu}$) for given initial data $H$ and $n_\mu\hat{g}^{\mu\nu}\partial_\nu H$ on $\Sigma$. But by the constraint equation $n_\mu E^\mu=0$ we have 
\begin{equation}
    0 = n_\mu E^{\mu}_{mhg} = n_\mu \hat{g}^{\mu\nu} \partial_\nu H
\end{equation}
and hence the initial data for $H$ is trivial and the unique solution is $H=0$. Similarly we can show $H^\mu = 0$ throughout $\hat{D}(\Sigma)$ using the Bianchi identity (\ref{eq:big}) and the constraint equation $n_\mu E^{\mu\nu}=0$, as explained in \cite{Kovacs:2020ywu}.

Therefore $E^\mu = 0$ and $E^{\mu\nu}=0$ in $\hat{D}(\Sigma)$. But since the causal cone of $g^{\mu\nu}$ lies inside that of $\hat{g}^{\mu\nu}$, we have $D(\Sigma)\subset \hat{D}(\Sigma)$ where $D(\Sigma)$ is the domain of dependence w.r.t. $g$. Hence, if the initial data satisfies the gauge conditions and constraints then,  within $D(\Sigma)$, 
a solution to the gauge fixed equations is also a solution to the equations of motion arising from the original Lagrangian.

\subsection{Principal Symbol}

The definition of the principal symbol is reviewed in the Appendix. The principal symbol for our equations acts on a vector of the form\footnote{Here the superscript ``T'' denotes a transpose, i.e., this is a column vector.}
\begin{equation}
    T_I=(t_{\mu\nu}, s_\rho)^T
\end{equation}
where $t_{\mu\nu}$ is symmetric. Indices $I, J, ...$ take values from 1 to 14. In geometric optics, $T_I$ describes the polarization of high frequency gravitoelectromagnetic waves. 

We label the blocks of the principal symbol\footnote{Here we are following the notation of \cite{Reall:2021}. ``g" stands for gravitational and ``m" stands for matter. In our case the matter is a Maxwell field.} as
\begin{equation}
    \mathcal{P}(\xi)^{IJ} = \mathcal{P}^{IJ\gamma\delta}\xi_\gamma \xi_\delta = \begin{pmatrix}
        \mathcal{P}_{gg}(\xi)^{\mu\nu\rho\sigma} & \mathcal{P}_{gm}(\xi)^{\mu\nu\rho} \\
        \mathcal{P}_{mg}(\xi)^{\mu\rho\sigma} & \mathcal{P}_{mm}(\xi)^{\mu\rho}
    \end{pmatrix}
\end{equation}
where we have suppressed the dependence on $(x,u,\partial_\mu u,\partial_0 \partial_i u, \partial_i \partial_j u)$. We decompose this into a part $\mathcal{P}_{\star}(\xi)$ coming from $E^\mu$ and $E^{\mu\nu}$, and a gauge fixing part $\mathcal{P}_{GF}(\xi)$ with
\begin{equation}
    \mathcal{P}(\xi)^{IJ} = \mathcal{P}_{\star}(\xi)^{IJ} + \mathcal{P}_{GF}(\xi)^{IJ} 
\end{equation}
\begin{equation}
    \mathcal{P}_{\star}(\xi)^{IJ} = \begin{pmatrix}
        \mathcal{P}_{gg\star}(\xi)^{\mu\nu\rho\sigma} & \mathcal{P}_{gm\star}(\xi)^{\mu\nu\rho} \\
        \mathcal{P}_{mg\star}(\xi)^{\mu\rho\sigma} & \mathcal{P}_{mm\star}(\xi)^{\mu\rho}
    \end{pmatrix}
\end{equation}
\begin{equation}
    \mathcal{P}_{GF}(\xi)^{IJ} = \begin{pmatrix}
        \hat{P}_{\alpha}^{\,\,\,\gamma\mu\nu} \tilde{P}^{\alpha\delta\rho\sigma} \xi_\gamma \xi_\delta & 0\\
        0 & -\hat{g}^{\mu\gamma}\tilde{g}^{\rho\delta}\xi_\gamma \xi_\delta
    \end{pmatrix}
\end{equation}
Furthermore, we split $\mathcal{P}_\star(\xi)$ into the standard Einstein-Maxwell terms (i.e. those coming from the first three terms in (\ref{action})) and the higher-derivative terms:
\begin{equation}
    \mathcal{P}_\star(\xi)^{IJ} = \mathcal{P}_{\star}^{EM}(\xi)^{IJ} + \delta\mathcal{P}_{\star}(\xi)^{IJ} 
\end{equation}
where
\begin{align}
    \mathcal{P}_{\star}^{EM}(\xi)^{IJ} &= \begin{pmatrix}
        (-\frac{1}{2}g^{\gamma\delta}P^{\mu\nu\rho\sigma}+P_{\alpha}^{\,\,\,\gamma\mu\nu}P^{\alpha\delta\rho\sigma})\xi_\gamma \xi_\delta & 0 \\
        0 & (-g^{\mu\rho}g^{\gamma\delta}+g^{\mu\gamma}g^{\rho\delta})\xi_\gamma \xi_\delta
    \end{pmatrix} \\
    &\equiv \begin{pmatrix}
        \mathcal{P}_{\star}^{E}(\xi)^{\mu\nu\rho\sigma} & 0 \\
        0 & \mathcal{P}_{\star}^{M}(\xi)^{\mu\rho}
    \end{pmatrix}
\end{align}
The expression for $\delta\mathcal{P}_{\star}(\xi)^{IJ}$ is lengthy and so is given in the Appendix in equation (\ref{deltaP}).

\subsection{Weak Coupling} \label{Weak}

The principal symbol is quadratic in $\xi_\mu$ so we can write e.g. ${\cal P}_\star(\xi)^{IJ} = P^{IJ\mu\nu}_\star \xi_\mu \xi_\nu$. We say that the theory is {\it weakly coupled} in some region of spacetime if a basis can be chosen in that region such that the components of  $\delta\mathcal{P}_{\star}^{IJ\mu\nu}$ are  small compared to the components of $(\mathcal{P}_{\star}^{EM})^{IJ\mu\nu}$. This is the condition that the contribution of the higher derivative terms to the principal symbol is small compared to the contribution from the 2-derivative terms. Note that this is a necessary condition for validity of EFT. 

We assume that the initial data is chosen so that the theory is weakly coupled on $\Sigma$.  By continuity, any solution arising from such data will remain weakly coupled at least for a small time. However, there is no guarantee that the solution will remain weakly coupled for all time e.g. weak coupling would fail if a curvature singularity forms. Under such circumstances, well-posedness may fail in the strongly coupled region but EFT would not be valid in this region anyway. 

\subsection{Symmetries of the Principal Symbol}

Our proof of strong hyperbolicity will make heavy use of the symmetries of the principal symbol. The following symmetries of the principal symbol are immediate from its definition:
\begin{gather}
    \mathcal{P}_{gg\star}(\xi)^{\mu\nu\rho\sigma} = \mathcal{P}_{gg\star}(\xi)^{(\mu\nu)\rho\sigma} = \mathcal{P}_{gg\star}(\xi)^{\mu\nu(\rho\sigma)} \label{eq:symgg}\\  \mathcal{P}_{gm\star}(\xi)^{\mu\nu\rho} = \mathcal{P}_{gm\star}(\xi)^{(\mu\nu)\rho} \label{eq:symgm}\\ \mathcal{P}_{mg\star}(\xi)^{\rho\mu\nu} = \mathcal{P}_{mg\star}(\xi)^{\rho(\mu\nu)} \label{eq:symmg}
\end{gather}
In \cite{Reall:2021}, it is shown that the fact $E^\mu$ and $E^{\mu\nu}$ are derived from an action principle leads to the following symmetries:
\begin{align}
    \mathcal{P}_{gg\star}(\xi)^{\mu\nu\rho\sigma} &= \mathcal{P}_{gg\star}(\xi)^{\rho\sigma\mu\nu}\\  \mathcal{P}_{gm\star}(\xi)^{\mu\nu\rho} &= \mathcal{P}_{mg\star}(\xi)^{\rho\mu\nu}\\  
    \mathcal{P}_{mm\star}(\xi)^{\mu\rho} &= \mathcal{P}_{mm\star}(\xi)^{\rho\mu}
\end{align}
In particular this means that $\mathcal{P}_{\star}(\xi)^{IJ}$ is symmetric.

The Bianchi identities (\ref{eq:big}) and (\ref{eq:bim}) together with the symmetries above also put conditions on the principal symbol (also given in \cite{Reall:2021} in an equivalent form), namely
\begin{align}
    \mathcal{P}_{gg\star}(\xi)^{\mu\nu\rho\sigma}\xi_\nu =\,&0 \label{eq:sgg}\\ 
    \mathcal{P}_{gm\star}(\xi)^{\mu\nu\rho}\xi_\nu =\,&0 \label{eq:sgm}\\  
    \mathcal{P}_{mg\star}(\xi)^{\mu\rho\sigma}\xi_\mu =\,&0 \label{eq:smg}\\  
    \mathcal{P}_{mm\star}(\xi)^{\mu\rho}\xi_\rho =\,&0  \label{eq:smm}
\end{align}

\section{Proof of Strong Hyperbolicity} \label{Proof}

\subsection{Characteristic equation}

\label{char_eq}

If $M(\xi_i)$ (defined by (\ref{Mmatrix})) is diagonalizable with real eigenvalues, and eigenvectors that depend smoothly on $\xi_i$ then a symmetrizer $K(\xi_i)$ can be defined by $K=(S^{-1})^\dagger S^{-1}$ where $S$ is the matrix whose columns are the eigenvectors. We will therefore start by considering the eigenvalue problem for $M(\xi_i)$, following the approach of \cite{Kovacs:2020ywu}. We will first prove smoothness of all eigenvectors for the modified harmonic gauge formulation standard 2-derivative Einstein-Maxwell, thus establishing the strong hyperbolicity of this formulation. We then consider the weakly coupled EFT. In this case we will not demonstrate smoothness of all eigenvectors but nevertheless we will explain how a symmetrizer can still be constructed. 

$M$ acts on vectors of the form\footnote{As before, the superscript ``T'' denotes a transpose, i.e., we are defining column vectors.} $v=(T_I, T'_I)^T$ where $T_I=(t_{\mu\nu}, s_\rho)^T$ and $T'_I=(t'_{\mu\nu}, s'_\rho)^T$. It is straightforward to show that any eigenvector of $M(\xi_i)$ with eigenvalue $\xi_0$ is of the form $(T_I, \xi_0 T_I)^T$ where $T_I$ satisfies
\begin{equation} \label{chareq}
    \mathcal{P}(\xi)^{IJ} T_J = 0
\end{equation}
where $\xi_\mu = (\xi_0, \xi_i)$. This is called the {\it characteristic equation}. If this equation admits a non-zero solution $T_I$ then $\xi_\mu$ is called a {\it characteristic covector} or simply characteristic. In geometric optics, characteristics arise as wavevectors of high frequency waves, with $T_I$ describing the polarization.

The characteristic equation can be rewritten as 
\begin{equation} \label{eq:eve}
    \begin{pmatrix}
        \mathcal{P}_{gg}(\xi)^{\mu\nu\rho\sigma}t_{\rho\sigma}+ \mathcal{P}_{gm}(\xi)^{\mu\nu\rho}s_\rho \\
        \mathcal{P}_{mg}(\xi)^{\mu\rho\sigma}t_{\rho\sigma} + \mathcal{P}_{mm}(\xi)^{\mu\rho}s_\rho
    \end{pmatrix} = \begin{pmatrix}
        0 \\
        0
    \end{pmatrix}
\end{equation}
We contract the first row of this equation with $\xi_\nu$ and the second row with $\xi_\mu$, and use (\ref{eq:sgg}-\ref{eq:smm}) to see that the non-gauge-fixing parts now vanish. After expanding the gauge fixing parts we get two equations:
\begin{align}
    -\frac{1}{2}\left(\hat{g}^{\nu\gamma}\xi_\nu\xi_\gamma\right)&\left(g^{\mu\beta} \tilde{P}_{\beta}^{\,\,\,\delta\rho\sigma} \xi_\delta t_{\rho\sigma} \right) = 0\\
    -\left(\hat{g}^{\mu\rho}\xi_\mu\xi_\rho\right)&\left(\tilde{g}^{\sigma\nu} \xi_\sigma s_\nu \right) = 0
\end{align}
Therefore we can split the analysis into two cases:
\renewcommand{\labelenumi}{\Roman{enumi}} 
\begin{enumerate}
    \item $\hat{g}^{\nu\gamma}\xi_\nu\xi_\gamma\neq  0 \Rightarrow g^{\mu\beta} \tilde{P}_{\beta}^{\,\,\,\delta\rho\sigma} \xi_\delta t_{\rho\sigma} = 0$ AND $\tilde{g}^{\sigma\nu} \xi_\sigma s_\nu = 0$
    \item $\hat{g}^{\nu\gamma}\xi_\nu\xi_\gamma = 0$
\end{enumerate}
Note that Case I implies that $\mathcal{P}_{GF}(\xi)^{IJ}T_J = 0$, i.e., $T_I$ ``satisfies the gauge conditions''. 

\subsection{Standard Einstein-Maxwell Theory} \label{Standard E-M}
We start our analysis with standard $c_3 = f = 0$ Einstein-Maxwell theory. Let's consider each case above.

\textbf{Case I}: This is defined by $\hat{g}^{\nu\gamma}\xi_\nu\xi_\gamma\neq  0$ which implies 
\begin{equation} \label{eq:gauge1}
    g^{\mu\beta} \tilde{P}_{\beta}^{\,\,\,\delta\rho\sigma} \xi_\delta t_{\rho\sigma} = 0
\end{equation}
and
\begin{equation} \label{eq:gauge2}
    \tilde{g}^{\sigma\nu} \xi_\sigma s_\nu = 0
\end{equation}
As noted above, these imply $\mathcal{P}_{GF}(\xi)^{IJ}T_J = 0$. Substituting this back into (\ref{chareq}) gives
\begin{equation}
    \mathcal{P}_\star(\xi)^{IJ} T_J = 0
\end{equation}
In standard E-M theory, $\mathcal{P}_\star(\xi) = \mathcal{P}_\star^{EM}(\xi)$ which is block diagonal and so this reduces to
\begin{equation} \label{eq:charEins}
    \mathcal{P}_{\star}^{E}(\xi)^{\mu\nu\rho\sigma}t_{\rho\sigma} = 0
\end{equation}
and
\begin{equation} \label{eq:charMax}
    \mathcal{P}_{\star}^{M}(\xi)^{\mu\rho} s_{\rho} = 0
\end{equation}
Hence we can use results derived in \cite{Kovacs:2020ywu} for the Einstein part, supplemented with their Maxwell equivalents. We split into two further cases:

\textbf{Subcase Ia}: This is defined by $g^{\gamma\delta}\xi_\gamma\xi_\delta\neq  0$. Expanding (\ref{eq:charMax}) gives 
\begin{equation}
    s^\mu = \xi^\mu \left(\frac{\xi^\rho s_\rho}{g^{\gamma\delta}\xi_\gamma\xi_\delta} \right) 
\end{equation}
Therefore $s_\rho = \lambda \xi_\rho$ for some $\lambda$. Substituting this into (\ref{eq:gauge2}) gives $\lambda \tilde{g}^{\sigma\nu} \xi_\sigma \xi_\nu = 0$. Similarly for the Einstein parts, \cite{Kovacs:2020ywu} shows that in the case $g^{\gamma\delta}\xi_\gamma\xi_\delta\neq  0$, the equations (\ref{eq:gauge1}) and (\ref{eq:charEins}) imply $t_{\mu\nu} = \xi_{(\mu}X_{\nu)}$ and $X^\mu \tilde{g}^{\sigma\nu} \xi_\sigma \xi_\nu = 0$ for some $X_\mu$. Taken together, for non-zero $T_I$ we have that 
\begin{equation} \label{eq:gaugexi}
    \tilde{g}^{\sigma\nu} \xi_\sigma \xi_\nu = 0
\end{equation}
and $T_I = (\xi_{(\mu}X_{\nu)}, \lambda \xi_\rho)^T$. Note that our assumption that the null cones of $\tilde{g}^{\mu\nu}$ and $g^{\mu\nu}$ do not intersect ensures that (\ref{eq:gaugexi}) is consistent with $g^{\gamma\delta}\xi_\gamma\xi_\delta\neq  0$.

Surfaces of constant $x^0$ are spacelike with respect to $g^{\mu\nu}$ and hence spacelike with respect to $\tilde{g}^{\mu\nu}$. Therefore, equation (\ref{eq:gaugexi}) has two real solutions $\tilde{\xi}^{\pm}_0$ that depend smoothly on $\xi_i$. The associated characteristic covectors are labelled $\tilde{\xi}^{\pm}_\mu = (\tilde{\xi}^{\pm}_0,\xi_i)$. The two solutions can be distinguished by the convention $\mp \tilde{g}^{0\nu}\tilde{\xi}^{\pm}_{\nu}>0$.

There are $4+1 = 5$ linearly independent eigenvectors associated to each eigenvalue $\tilde{\xi}^{\pm}_0$, given by the arbitrary choices of $X_\mu$ and $\lambda$. We call these "pure gauge'' eigenvectors because they arise from a 4-dimensional residual gauge freedom in $g_{\mu\nu}$ and a 1-dimensional residual gauge freedom in $A_\mu$ in the gauge fixed equations of motion (\ref{eq:gfg}) and (\ref{eq:gfm}). These eigenvectors form $5$-dimensional eigenspaces, which we denote as $\tilde{V}^\pm$.

\textbf{Subcase Ib}: This is defined by 
\begin{equation} \label{eq:physxi}
    g^{\sigma\nu} \xi_\sigma \xi_\nu = 0
\end{equation}
Again this has two real solutions $\xi^{\pm}_0$ with $\mp g^{0\nu}\xi^{\pm}_{\nu}>0$ and characteristic covector $\xi^{\pm}_\mu = (\xi^{\pm}_0,\xi_i)$. We will find the dimension of the space of eigenvectors. Starting with the Maxwell part, the equation $\mathcal{P}_{\star}^{M}(\xi^{\pm}_\mu)^{\mu\rho} s_{\rho} = 0$ reduces to 
\begin{equation} \label{eq:caseib}
    g^{\sigma\nu} \xi^{\pm}_\sigma s_\nu = 0
\end{equation}
But the Case I condition (\ref{eq:gauge2}) is
\begin{equation} \label{eq:caseib2}
    \tilde{g}^{\sigma\nu} \xi^{\pm}_\sigma s_\nu = 0
\end{equation}
Hence the only requirements on the polarization $s_\rho$ are that it is orthogonal to $\xi^{\pm}_\mu$ with respect to both $g^{\mu\nu}$ and $\tilde{g}^{\mu\nu}$. These are the "physical" photon polarizations, and for each eigenvalue $\xi^{\pm}_0$, the corresponding $T_I = (0,s_\rho)$ form a 2-dimensional eigenspace that depends smoothly on $\xi_i$.

For the metric part, \cite{Kovacs:2020ywu} proves a similar statement in this case. For each eigenvalue $\xi^{\pm}_0$, there are 2 linearly independent eigenvectors with $T_I = (t_{\mu\nu},0)$ that depend smoothly on $\xi_i$. These polarizations are transverse with respect to $g^{\mu\nu}$ and $\tilde{g}^{\mu\nu}$ in the sense that $P_{\beta}^{\,\,\,\delta\rho\sigma} \xi^{\pm}_\delta t_{\rho\sigma} = 0$ and $\tilde{P}_{\beta}^{\,\,\,\delta\rho\sigma} \xi^{\pm}_\delta t_{\rho\sigma} = 0$, and correspond to physical polarizations of the metric.

Therefore there is a 4-dimensional eigenspace $V^{\pm}$ for each eigenvalue $\xi^{\pm}_0$, with eigenvectors depending smoothly on $\xi_i$.

\textbf{Case II}: This is defined by 
\begin{equation} \label{eq:violxi}
    \hat{g}^{\sigma\nu} \xi_\sigma \xi_\nu = 0
\end{equation}
Once again, this has two real solutions $\hat{\xi}^{\pm}_0$ which we distinguish by $\mp \hat{g}^{0\nu}\hat{\xi}^{\pm}_{\nu}>0$. Since the characteristic covectors $\hat{\xi}^{\pm}_\mu = (\hat{\xi}^{\pm}_0, \xi_i)$ are the same as those for (\ref{eq:divbianchi}), we call the corresponding eigenvectors "gauge condition-violating". (In geometric optics these correspond to high frequency solutions of the gauge-fixed equations that violate the gauge condition.)

We first look at $\hat{\xi}^{+}_0$ and construct its eigenvectors. Since $\hat{\xi}^{+}_\mu$ is null with respect to $\hat{g}^{\mu\nu}$, it is spacelike with respect to $g^{\mu\nu}$. Therefore we can introduce a basis $\{e_0^\mu,e_1^\mu,e_2^\mu,e_3^\mu\}$ which is orthonormal with respect to $g_{\mu\nu}$ and $e_1^\mu \propto \hat{\xi}^{+\mu}$ (recall that indices are raised using $g^{\mu\nu}$). This basis can be chosen to depend smoothly on $\xi_i$ \cite{Kovacs:2020ywu}. 

Define indices $A, B, ...$ to take values 0, 2, 3. In this basis we can write a general symmetric tensor as
\begin{equation} \label{eq:tmu}
    t_{\mu\nu} = \hat{\xi}^{+}_{(\mu} X_{\nu)} + t_{AB} e_\mu^A e_\nu^B
\end{equation}
and a general covector as 
\begin{equation} \label{eq:srho}
    s_\rho = \lambda\hat{\xi}^{+}_\rho + s_C e_\rho^C
\end{equation}
By the conditions (\ref{eq:sgg}-\ref{eq:smm}) and symmetries (\ref{eq:symgg}-\ref{eq:symmg}) the only non-vanishing components of $\mathcal{P}_{\star}(\hat{\xi}^{+})$ are those with $A, B, ...$ indices. To construct the eigenvectors, we start by considering solutions $(t_{AB}, s_C)$ to the following
\begin{equation} \label{eq:mat}
    \begin{pmatrix}
        \mathcal{P}^E_{\star}(\hat{\xi}^{+})^{ABCD} & 0 \\
        0 & \mathcal{P}^M_{\star}(\hat{\xi}^{+})^{AC}
    \end{pmatrix}
    \begin{pmatrix}
        t_{CD} \\
        s_C
    \end{pmatrix} = \begin{pmatrix}
        \hat{P}_{\alpha}^{\,\,\,\beta AB} \hat{\xi}^{+}_{\beta} v^{\alpha} \\
        \hat{g}^{A\alpha}\hat{\xi}^{+}_{\alpha}w
    \end{pmatrix}
\end{equation}
where $(v^\alpha,w)$ is a fixed constant vector. We claim this can be solved uniquely. Consider an element $(r_{AB}, p_C)$ of the kernel of the matrix on the left hand side:
\begin{align}
    \mathcal{P}_{\star}^{E}(\hat{\xi}^{+})^{ABCD} r_{CD} &= 0 \\
    \mathcal{P}_{\star}^{M}(\hat{\xi}^{+})^{A C} p_C = 0 \label{eq:ker}
\end{align}
In \cite{Kovacs:2020ywu} it is shown that $\mathcal{P}_{\star}^{E}(\hat{\xi}^{+})^{ABCD}$ has trivial kernel and so $r_{AB}=0$. We can do the same for $\mathcal{P}_{\star M}(\hat{\xi}^{+})^{A B}$ using a similar argument: (\ref{eq:ker}) implies that 
\begin{equation}
     \mathcal{P}_{\star}^{M}(\hat{\xi}^{+})^{\mu\nu} p_\nu = 0
\end{equation}
for any $p_1$. Expanding implies
\begin{equation}
    p^\mu (g^{\gamma\delta}\hat{\xi}^{+}_\gamma \hat{\xi}^{+}_\delta) = \hat{\xi}^{+\mu} (g^{\gamma\delta}\hat{\xi}^{+}_\gamma p_\delta)
\end{equation}
The null cones of $g^{\gamma\delta}$ and $\hat{g}^{\gamma\delta}$ do not intersect and so $g^{\gamma\delta}\hat{\xi}^{+}_\gamma \hat{\xi}^{+}_\delta \neq 0$. This means the above implies $p^\mu \propto \hat{\xi}^{+\mu}$, and hence in our orthonormal basis, $p_A = 0$, which establishes the result.

Therefore the kernel of the matrix on the left hand side of (\ref{eq:mat}) is trivial and so there is a unique solution $(t_{AB}(v^\alpha,w), s_C(v^\alpha,w))$ to (\ref{eq:mat}). This solution depends smoothly on $(v^\alpha, w)$, $\xi_i$ and $g_{\mu\nu}$ since both sides of (\ref{eq:mat}) depend smoothly on these things.

We use these values for $(t_{AB}(v^\alpha,w), s_C(v^\alpha,w))$ in our definitions of $(t_{\mu\nu},s_\rho)$ defined by (\ref{eq:tmu}) and (\ref{eq:srho}). This implies that 
\begin{equation} \label{eq:fix}
    \mathcal{P}_{\star}(\hat{\xi}^{+})^{IJ}  \begin{pmatrix}
        t_{\rho\sigma}(v^\gamma,w) \\
        s_\rho (v^\gamma,w)
    \end{pmatrix} = \begin{pmatrix}
        \hat{P}_{\alpha}^{\,\,\,\beta\mu\nu} \hat{\xi}^{+}_{\beta} v^{\alpha} \\
        \hat{g}^{\mu\alpha}\hat{\xi}^{+}_{\alpha}w
    \end{pmatrix} 
\end{equation}
since the components with 1-indices vanish on the top line and bottom lines because both sides have vanishing contractions with $\hat{\xi}^{+}_\mu$.

Now, in our definitions (\ref{eq:tmu}) and (\ref{eq:srho}) we judiciously chose
\begin{equation}
    X^\mu(v,t_{AB}) = \frac{2}{\tilde{g}^{\gamma\delta}\hat{\xi}^{+}_\gamma \hat{\xi}^{+}_\delta}\left(v^\mu - \tilde{P}^{\mu\nu AB} \hat{\xi}^{+}_{\nu} t_{AB}\right)
\end{equation}
and
\begin{equation}
    \lambda(w,s_C) = \frac{1}{\tilde{g}^{\gamma\delta}\hat{\xi}^{+}_\gamma \hat{\xi}^{+}_\delta}\left(w-\tilde{g}^{\gamma A}\hat{\xi}^{+}_\gamma s_A\right)
\end{equation}
which also depend smoothly on their arguments (and $\xi_i$). These choices imply
\begin{align}
    \tilde{P}^{\mu\nu\rho\sigma}\hat{\xi}^{+}_{\nu} t_{\rho\sigma} &= \frac{1}{2}\left(\tilde{g}^{\gamma\delta}\hat{\xi}^{+}_\gamma \hat{\xi}^{+}_\delta\right)X^\mu + \tilde{P}^{\mu\nu AB} \hat{\xi}^{+}_{\nu} t_{AB} \\ \nonumber
    &= v^\mu - \tilde{P}^{\mu\nu AB} \hat{\xi}^{+}_{\nu} t_{AB} + \tilde{P}^{\mu\nu AB} \hat{\xi}^{+}_{\nu} t_{AB} \\ \nonumber
    &= v^\mu \label{eq:vmu}
\end{align}
and 
\begin{align}
    \tilde{g}^{\gamma\nu}\hat{\xi}^{+}_\gamma s_\nu &= \lambda \tilde{g}^{\gamma\nu}\hat{\xi}^{+}_\gamma \hat{\xi}^{+}_\nu + \tilde{g}^{\gamma A}\hat{\xi}^{+}_\gamma s_A \\ \nonumber
    &= w \label{eq:w}
\end{align}
which imply that $T_I = (t_{\mu\nu}(v^\alpha, w), s_\rho(v\alpha, w)$ satisfies
\begin{align}
    \mathcal{P}(\hat{\xi}^{+})^{IJ} T_J &= \mathcal{P}_\star(\hat{\xi}^{+})^{IJ} T_J + \mathcal{P}_{GF}(\hat{\xi}^{+})^{IJ} T_J \\
    &= \begin{pmatrix}
        \hat{P}_{\alpha}^{\,\,\,\beta\mu\nu} \hat{\xi}^{+}_{\beta} v^{\alpha} \\
        \hat{g}^{\mu\alpha}\hat{\xi}^{+}_{\alpha}w
    \end{pmatrix} - \begin{pmatrix}
        \hat{P}_{\alpha}^{\,\,\,\gamma\mu\nu} \hat{\xi}^{+}_\gamma \tilde{P}^{\alpha\delta\rho\sigma} \hat{\xi}^{+}_\delta t_{\rho\sigma} \\
        \hat{g}^{\mu\gamma}\tilde{g}^{\nu\delta}\hat{\xi}^{+}_\gamma \hat{\xi}^{+}_\delta s_\nu
    \end{pmatrix} \nonumber\\
    &= \begin{pmatrix}
        \hat{P}_{\alpha}^{\,\,\,\beta\mu\nu} \hat{\xi}^{+}_{\beta} v^{\alpha} \\
        \hat{g}^{\mu\alpha}\hat{\xi}^{+}_{\alpha}w
    \end{pmatrix} - \begin{pmatrix}
        \hat{P}_{\alpha}^{\,\,\,\beta\mu\nu} \hat{\xi}^{+}_{\beta} v^{\alpha} \\
        \hat{g}^{\mu\alpha}\hat{\xi}^{+}_{\alpha}w
    \end{pmatrix} \nonumber\\
    &= 0 \nonumber
\end{align}
where the second equality comes from (\ref{eq:fix}) and the third equality comes from (\ref{eq:vmu}) and (\ref{eq:w}). 

Hence for every $(v^\alpha,w)$ we have constructed a smoothly varying eigenvector $(T_I(v^\alpha,w), \hat{\xi}^{+}_0 T_I(v^\alpha,w))$ of $M$ with eigenvalue $\hat{\xi}^{+}_0$. If we pick a set of 5 linearly independent choices of $(v^\alpha, w)$ then the corresponding $t_{AB}$ and $s_C$ will be linearly independent by the triviality of the kernel of the LHS of (\ref{eq:mat}), and hence the corresponding eigenvectors will be linearly independent. Label the 5-dimensional span of these eigenvectors by $\hat{V}^{+}$. We can repeat all the above steps with $\hat{\xi}_0^{-}$ to get the same result for $\hat{V}^{-}$. We claim that $\hat{V}^{\pm}$ contain all the eigenvectors with eigenvalue $\hat{\xi}_0^{\pm}$ by counting the dimensions of the eigenspaces we have found so far:
\begin{equation}
    \dim \tilde{V}^+ + \dim \tilde{V}^- + \dim V^+ + \dim V^- + \dim \hat{V}^+ + \dim \hat{V}^- = 5 + 5 + 4 + 4 + 5 + 5 = 28
\end{equation}
$M$ is a 28 by 28 matrix, and hence there are no more eigenvectors to find. Therefore $\hat{V}^{\pm}$ are the total eigenspaces for $\hat{\xi}_0^{\pm}$.

To summarise, we have found that, for standard 2-derivative Einstein-Maxwell, $M(\xi)$ has 6 distinct eigenvalues, $\tilde{\xi}_0^{\pm}$, $\xi_0^{\pm}$ and $\hat{\xi}_0^{\pm}$ which are all real. Furthermore it has a complete set of eigenvectors that depend smoothly on $\xi_i$. Therefore the modified-harmonic-gauge formulation of standard Einstein-Maxwell is strongly hyperbolic.

Now, as mentioned in Section \ref{Strong Hyperbolicity}, the argument that strong hyperbolicity implies well-posedness assumes that $M$ is invertible. This is equivalent to $C^{IJ}$ being invertible, which is equivalent to the condition that $\xi_0 \neq 0$ for any characteristic covector. However, we chose our spacetime foliation such that surfaces of constant $x^0$ are spacelike with respect to $g^{\mu\nu}$, and hence spacelike with respect to $\tilde{g}^{\mu\nu}$ and $\hat{g}^{\mu\nu}$. This means that a covector with $\xi_0 = 0$ is spacelike with respect to all three (inverse) metrics. But as we found above, the characteristic covectors are null with respect to one of the three metrics, and hence $M$ is invertible for standard E-M theory. By continuity, $M$ will remain invertible when we include higher derivative terms, assuming weak coupling. 

\subsection{Weakly Coupled Einstein-Maxwell EFT}

Now we consider our theory including the higher derivative terms. 
At weak coupling, $M(\xi_i)$ is a small deformation of the corresponding matrix of the 2-derivative theory. The continuity of this deformation will be used in the following to show that many of the above results still hold. In particular, the eigenvalues of $M(\xi_i)$ will be close to those discussed above, and can be sorted into 6 groups corresponding to which eigenvalue they approach in the standard 2-derivative (i.e. $c_3\rightarrow0$, $f\rightarrow0$) limit. As in \cite{Kovacs:2020ywu}, we call these groups the $\tilde{\xi}_0^{+}$-group, the $\tilde{\xi}_0^{-}$-group etc. We don't know that these eigenvalues are real so we view $M(\xi_i)$ as acting on the 28 dimensional space $V$ of {\it complex} vectors of the form $v=(T_I, T'_I)^T$. 

Following the argument of \cite{Kovacs:2020ywu} we can decompose $V$ as follows
\begin{equation}
    V = \tilde{V}^{+} \oplus \hat{V}^{+} \oplus V^{+} \oplus \tilde{V}^{-} \oplus \hat{V}^{-} \oplus V^{-}
\end{equation}
where $V^+$ is the sum of all generalized eigenspaces\footnote{The generalized eigenspace corresponding to a matrix $A$ and eigenvalue $\lambda$ is the space of vectors $x$ such that there exists a positive integer $m$ with $(A-\lambda I)^m x = 0$. In the Jordan decomposition of $A$, each Jordan block is associated with a generalized eigenspace.} associated with eigenvalues in the $\xi_0^+$ group and similarly for the other spaces. The spaces $\tilde{V}^{+}$, $\tilde{V}^{-}$ etc. must have the same dimensions as the corresponding eigenspaces in standard 2-derivative E-M theory. Therefore, $\tilde{V}^{\pm}$ and $\hat{V}^{\pm}$ are all five dimensional and $V^{\pm}$ are four dimensional. These vector spaces are complex. 

Recall (section \ref{char_eq}) that the analysis of the characteristic equation splits into two cases. We will see that $\tilde{V}^{\pm}$ correspond to eigenvectors in Case I arising from the same residual gauge invariance as in the standard 2-derivative theory. $V^{\pm}$ correspond to the remaining eigenvectors in Case I, which are the physical eigenvectors. $\hat{V}^{\pm}$ correspond to gauge-condition-violating eigenvectors in Case II. 
\subsection{$\tilde{V}^{\pm}$}

These are the spaces associated with the $\tilde{\xi}^{\pm}_0$-groups of eigenvalues, where $\tilde{\xi}^{\pm}_0$ are the two real solutions to $\tilde{g}^{\mu\nu}\tilde{\xi}^{\pm}_\mu\tilde{\xi}^{\pm}_\nu = 0$. However, the weakly coupled theory still has the same residual gauge freedoms in $g_{\mu\nu}$ and $A_\mu$ as the 2-derivative theory. As such it turns out that $\tilde{\xi}^{\pm}_0$ are still eigenvalues of the weakly coupled theory with the same eigenvectors $v=(T_I, \xi_0 T_I)^T$ of the form
\begin{equation}
    T_I = (\tilde{\xi}^{\pm}_{(\mu} X_{\nu)}, \lambda \tilde{\xi}^{\pm}_\rho)
\end{equation}
for arbitrary $X_\nu$ and $\lambda$. To show this, note that $T_I$ satisfies $\mathcal{P}_{GF}(\tilde{\xi}^{\pm})^{IJ}T_J = 0$, and also 
\begin{align}
    \mathcal{P}_{\star}(\tilde{\xi}^{\pm})^{IJ} T_J &= \begin{pmatrix}
        \mathcal{P}_{gg\star}(\tilde{\xi}^{\pm})^{\mu\nu\rho\sigma}\tilde{\xi}^{\pm}_{(\rho} X_{\sigma)}+ \lambda \mathcal{P}_{gm\star}(\tilde{\xi}^{\pm})^{\mu\nu\rho} \tilde{\xi}^{\pm}_\rho \\
        \mathcal{P}_{mg\star}(\tilde{\xi}^{\pm})^{\mu\rho\sigma}\tilde{\xi}^{\pm}_{(\rho} X_{\sigma)} + \lambda\mathcal{P}_{mm\star}(\tilde{\xi}^{\pm})^{\mu\rho}\tilde{\xi}^{\pm}_\rho 
    \end{pmatrix} 
    = \begin{pmatrix}
        0 \\
        0
    \end{pmatrix}
\end{align}
where the second equality follows from conditions (\ref{eq:sgg}-\ref{eq:smm}) and symmetries (\ref{eq:symgg}) and (\ref{eq:symmg}). Therefore $\mathcal{P}(\tilde{\xi}^{\pm})^{IJ}T_J = 0$. Hence $\tilde{V}^{\pm}$ are genuine eigenspaces (rather than generalized eigenspaces) with eigenvalues $\tilde{\xi}^{\pm}_0$, and eigenvectors that depend smoothly on $\xi_i$. 

\subsection{$\hat{V}^{\pm}$}

These are the spaces associated with the $\hat{\xi}^{\pm}_0$-groups of eigenvalues where $\hat{\xi}^{\pm}_0$ are the two real solutions to $\hat{g}^{\mu\nu}\hat{\xi}^{\pm}_\mu\hat{\xi}^{\pm}_\nu = 0$. We'll now show that our above construction of eigenvectors with eigenvalue $\hat{\xi}^{\pm}_0$ in the 2-derivative theory extends to the weakly coupled higher-derivative theory with only minor modifications.

We introduce the same basis as before, and take indices A,B,... to take values 0,2,3. The only line which needs changing is equation (\ref{eq:mat}), as we now want to find solutions $(t_{AB}, s_C)$ to

\begin{equation} \label{eq:matmod}
    \begin{pmatrix}
        \mathcal{P}_{gg\star}(\hat{\xi}^{+})^{ABCD} & \mathcal{P}_{gm\star}(\hat{\xi}^{+})^{ABC} \\
        \mathcal{P}_{mg\star}(\hat{\xi}^{+})^{ACD} & \mathcal{P}_{mm\star}(\hat{\xi}^{+})^{AC}
    \end{pmatrix}
    \begin{pmatrix}
        t_{CD} \\
        s_C
    \end{pmatrix} = \begin{pmatrix}
        \hat{P}_{\alpha}^{\,\,\,\beta AB} \hat{\xi}^{+}_{\beta} v^{\alpha} \\
        \hat{g}^{A\alpha}\hat{\xi}^{+}_{\alpha}w
    \end{pmatrix}
\end{equation}
However in standard E-M theory, we found that the kernel of the matrix on the left hand side is trivial and so its determinant is non-zero. By continuity, its determinant is also non-zero for sufficiently weak coupling, and so its kernel is still trivial and there is still a unique solution $(t_{AB}(v^\alpha,w), s_C(v^\alpha,w))$ to (\ref{eq:matmod}). Both sides of (\ref{eq:matmod}) still depend smoothly on $(v^\alpha, w)$, $\xi_i$, $g_{\mu\nu}$ and $A_\mu$ and their derivatives, and so the solution also depends smoothly on these things.

The rest of the construction follows the same steps as for the 2-derivative theory, and hence $\hat{V}^{\pm}$ are genuine eigenspaces with eigenvalues $\hat{\xi}^{\pm}_0$ and eigenvectors that depend smoothly on $\xi_i$ and the fields and their derivatives.

\subsection{$V^{\pm}$}

These are the spaces associated with the $\xi^{\pm}_0$-groups of eigenvalues where $\xi^{\pm}_0$ are the two real solutions to $g^{\mu\nu}\xi^{\pm}_\mu\xi^{\pm}_\nu = 0$. Since we are only considering weak coupling, we can assume that these eigenvalues are sufficiently close to $\xi^\pm_0$ so that 
 $\hat{g}^{\sigma\nu}\xi_\sigma \xi_\nu \neq 0$ and $\tilde{g}^{\sigma\nu}\xi_\sigma \xi_\nu \neq 0$. Therefore the eigenvalues and eigenvectors in the 4-dimensional generalized eigenspaces $V^{\pm}$ are those in Case I that don't also satisfy $\tilde{g}^{\sigma\nu}\xi_\sigma \xi_\nu = 0$. We will show that $V^{\pm}$ are genuine eigenspaces by closely following the argument in \cite{Kovacs:2020ywu} for Horndeski theories.

The first step of the argument is to establish that the deformed eigenvalues are real. We proceed by defining
\begin{equation}
    H^{\pm}_\star = \pm
    \begin{pmatrix}
        B_\star & A_\star \\
        A_\star & 0 
    \end{pmatrix}
\end{equation}
where $A_\star$ and $B_\star$ are defined as in (\ref{eq:ABCdef}) but by only using the non-gauge fixing parts of the principal symbol. $H^{\pm}_\star$ is Hermitian since $\mathcal{P}_{\star}(\xi)^{IJ}$ is symmetric and real. We then define the Hermitian form $(,)_\pm$ on vectors $v^{(i)} = (T^{(i)}_I, T'^{(i)}_I)$ in $V^\pm$ (viewed as 4-dimensional complex vector spaces) by
\begin{equation} \label{eq:8}
    (v^{(1)},v^{(2)})_\pm = v^{(1)\dagger} H^{\pm}_\star v^{(2)}
\end{equation}
We show this is positive definite for standard E-M theory, and hence by continuity it is positive definite for sufficiently weakly coupled E-M theory. In standard E-M, $V^\pm$ are genuine eigenspaces each with one eigenvalue $\xi^{\pm}_0$ and eigenvectors $v^{(i)} = (t_{\mu\nu}^{(i)},s_\rho^{(i)},\xi^{\pm}_0 t_{\mu\nu}^{(i)},\xi^{\pm}_0 s_\rho^{(i)})$ satisfying the equations which define Subcase Ib. $A_\star$ and $B_\star$ are also block diagonal so the Hermitian form splits into a gravitational part and a Maxwell part:
\begin{equation}
    (v^{(1)},v^{(2)})_\pm = \pm \left[ t^{(1)\ast}_{\mu\nu}\left(2\xi^{\pm}_0 A_\star + B_\star\right)^{\mu\nu\rho\sigma} t^{(2)}_{\rho \sigma} + s^{(1)\ast}_{\mu}\left(2\xi^{\pm}_0 A_\star + B_\star\right)^{\mu\rho} s^{(2)}_{\rho} \right]
\end{equation}
In \cite{Kovacs:2020ywu} it is shown that the gravitational part simplifies to $-g^{0\nu}\xi^{\pm}_\nu t^{(1)\ast}_{\mu\nu}P^{\mu\nu\rho\sigma}t^{(2)}_{\rho \sigma}$ where
\be
P_{\alpha}^{\,\,\,\beta\mu\nu} = \delta_\alpha^{(\mu}g^{\nu)\beta}-\frac{1}{2}\delta_\alpha^\beta g^{\mu\nu}
\ee
Using (\ref{eq:caseib}), we can also reduce the Maxwell part, leading to
\begin{equation}
    (v^{(1)},v^{(2)})_\pm = \mp g^{0\nu}\xi^{\pm}_\nu\left[ t^{(1)\ast}_{\mu\nu}P^{\mu\nu\rho\sigma}t^{(2)}_{\rho \sigma} + 2 s^{(1)\ast}_{\mu}g^{\mu\rho} s^{(2)}_{\rho} \right]
\end{equation}
To simplify further, we pick a tangent space basis $(e_0)^\mu = \xi^{\pm\mu}$, $(e_1)^\mu \propto \xi^{\mp\mu}$, $(e_2)^\mu$ and $(e_3)^\mu$ with
\begin{equation}
    g(e_0,e_1) = 1, \,\, g(e_{\hat{i}},e_{\hat{j}}) = \delta_{\hat{i}\hat{j}}
\end{equation}
where $\hat{i}, \hat{j} = 2,3$ and all other contractions vanish. In this basis, equation (\ref{eq:caseib}) becomes $s_0 = 0$, and so $s^{(1)\ast}_{\mu}g^{\mu\rho} s^{(2)}_{\rho} = s^{(1)\ast}_{\hat{i}}s^{(2)}_{\hat{i}}$. Similarly for the gravitational part, \cite{Kovacs:2020ywu} shows that the conditions defining Subcase Ib imply that all components of $t_{\mu\nu}$ either vanish or depend linearly on the traceless quantity $t_{\hat{i}\hat{j}}$. Furthermore, they show that $t^{(1)\ast}_{\mu\nu}P^{\mu\nu\rho\sigma}t^{(2)}_{\rho \sigma} = t^{(1)\ast}_{\hat{i}\hat{j}}t^{(2)}_{\hat{i}\hat{j}}$. Therefore
\begin{equation}
    (v^{(1)},v^{(2)})_\pm = \mp \xi^{\pm0}\left( t_{\hat{i}\hat{j}}^{(1)\ast}\, t_{\hat{i}\hat{j}}^{(2)}+2s_{\hat{i}}^{(1)\ast}\, s_{\hat{i}}^{(2)}\right)
\end{equation}
Our convention was that $\mp \xi^{\pm0} > 0$ so this is non-negative for $v^{(1)}=v^{(2)}=v$. Suppose that $(v,v)_\pm=0$. Then $s_{\hat{i}}=0$ and so equation (\ref{eq:caseib2}) becomes $\tilde{g}^{11}s_1 = 0$. But $0\neq \tilde{g}^{\sigma\nu}\xi^{\pm}_\sigma \xi^{\pm}_\nu = \tilde{g}^{11}$, and hence this implies $s_1 = 0$. Therefore $s_\rho = 0$. Similarly $t_{\hat{i}\hat{j}}=0$ implies that $t_{\mu\nu} = 0$ by Subcase Ib conditions \cite{Kovacs:2020ywu}. Hence $(v,v)_\pm\ge0$ with equality iff $v=0$. Therefore we have shown that the Hermitian form is positive definite for $2$-derivative E-M theory, and hence, by continuity, also for weakly coupled E-M theory.

In \cite{Kovacs:2020ywu}, it is shown that the existence of this positive definite form on the complex vector space $V^\pm$ implies the eigenvalues in the $\xi_0^\pm$-group are real so long as a) $\mathcal{P}_{\star}(\xi)^{IJ}$ is symmetric (which we have) and b) the corresponding eigenvectors are in the kernel of $\mathcal{P}_{GF}(\xi)^{IJ}$. The second condition follows for $V^{\pm}$ by Case I conditions. Hence the eigenvalues in the $\xi_0^\pm$-groups are real.

We now proceed to show diagonalisability of $M(\xi_i)$. Since we've already shown that $\tilde{V}^\pm$ and $\hat{V}^\pm$ are genuine eigenspaces, we just need to show that $M(\xi_i)$ is diagonalizeable within $V^\pm$. To this end, let $\xi_0$ be an eigenvalue in the $\xi_0^\pm$ group and consider a {\it left} eigenvector of $M(\xi_i)$ with this eigenvalue. One can show these are of the form
\begin{equation}
    w = \left(T_I, \xi_0 T_I\right) \begin{pmatrix}
        B & A \\
        A & 0
    \end{pmatrix}
\end{equation}
where
\begin{equation}
    T_I \mathcal{P}(\xi)^{IJ} = 0
\end{equation}
Now, using the symmetries of $\mathcal{P}_{\star}(\xi)^{IJ}$ one can show that a family of left eigenvectors with eigenvalue $\hat{\xi}_0^{\pm}$ is given by 
\begin{equation} \label{eq:fam}
    T_I = (\hat{\xi}_{(\mu}^{\pm}X_{\nu)}, \lambda \hat{\xi}_\rho^{\pm})
\end{equation}
for arbitrary $X_\nu$ and $\lambda$. Then by considering the Jordan normal form of $M$, we have that $w$ must be orthogonal to any vector $v=(u_I,u'_I)^T$ in any of $V^{\pm}$ or $\tilde{V}^{\pm}$, i.e.
\begin{align}
    0 = wv &= \left(T_I, \hat{\xi}_0^{\pm} T_I\right) \begin{pmatrix}
        B^{IJ} & A^{IJ} \\
        A^{IJ} & 0
    \end{pmatrix} \begin{pmatrix}
        u_J \\
        u'_J
    \end{pmatrix} \\
    &= T_I(B^{IJ}+ \hat{\xi}_0^{\pm} A^{IJ})u_J + T_I A^{IJ} u'_J
\end{align}
By expanding $T_I$ through (\ref{eq:fam}) and using the fact that $X_\nu$ and $\lambda$ are arbitrary we get the two following conditions on $(u_I, u'_I)$:
\begin{gather}
    0 = \hat{\xi}_\nu^{\pm} (B^{\mu\nu J} + \hat{\xi}_0^{\pm} A^{\mu\nu J})u_J + \hat{\xi}_\nu^{\pm} A^{\mu\nu J} u'_J \\
    0 = \hat{\xi}_\mu^{\pm} (B^{\mu J} + \hat{\xi}_0^{\pm} A^{\mu J})u_J + \hat{\xi}_\mu^{\pm} A^{\mu J} u'_J
\end{gather}
We can eliminate $(\hat{\xi}_0^{\pm})^2$ using $\hat{g}^{\sigma\nu}\hat{\xi}_\sigma \hat{\xi}_\nu = 0$ to get
\begin{gather}
    \hat{\xi}_0^{\pm}R^\mu +S^\mu = 0 \label{eq:vec} \\ 
    \hat{\xi}_0^{\pm}R +S = 0 \label{eq:sca}
\end{gather}
where
\begin{gather}
    R^\mu \equiv -2(\hat{g}^{00})^{-1}\hat{g}^{0i}\xi_i A^{\mu 0J}u_J + B^{\mu 0 J}u_J + \xi_i A^{\mu i J}u_J + A^{\mu 0 J}u'_J \label{eq:Rmu} \\
    R \equiv -2(\hat{g}^{00})^{-1}\hat{g}^{0i}\xi_i A^{0J}u_J + B^{0 J}u_J + \xi_i A^{i J}u_J + A^{0 J}u'_J \label{eq:R} \\
    S^\mu \equiv -(\hat{g}^{00})^{-1}\hat{g}^{ij}\xi_i\xi_j A^{\mu 0 J}u_J+\xi_i B^{\mu i J}u_J + \xi_i A^{\mu i J}u'_J \\
    S \equiv -(\hat{g}^{00})^{-1}\hat{g}^{ij}\xi_i\xi_j A^{0 J}u_J+\xi_i B^{i J}u_J + \xi_i A^{i J}u'_J
\end{gather}
Note that none of $R^\mu$, $R$, $S^\mu$ or $S$ depend on $\hat{\xi}^{\pm}_0$. Therefore, since (\ref{eq:vec}) and (\ref{eq:sca}) are true for both signs $\pm$, they imply
\begin{equation}
    R^\mu = 0 = S^\mu,\,\,\,\,\,\, R = 0 = S
\end{equation}
Now, we can match co-efficients of powers of $\xi_0$ in the Bianchi-style identities (\ref{eq:sgg}-\ref{eq:smm}) to get the following identities on $A_\star$ and $B_\star$ (and on $C_\star$ but these aren't relevant to our argument):
\begin{align}
    A_\star^{\mu 0 I} &= 0 \\
    \xi_i A_\star^{\mu i I} + B_\star^{\mu 0 I} &= 0 \\
    A_\star^{0 I} &= 0 \\
    \xi_i A_\star^{i I} + B_\star^{0 I} &= 0
\end{align}
We can plug these into (\ref{eq:Rmu}) and (\ref{eq:R}) and see that all the $\star$ terms vanish in $R^\mu$ and $R$. Hence they only depend on gauge-fixing terms which are block diagonal. Writing $u_I = (t_{\mu\nu}, s_\rho)$ and $u'_I = (t'_{\mu\nu}, s'_\rho)$ and expanding $A_{GF}$ and $B_{GF}$, the equations $R^\mu=0$ and $R=0$ reduce to
\begin{gather}
    \tilde{P}_{\beta}^{\,\,\,i\rho\sigma} \xi_i t_{\rho\sigma} + \tilde{P}_{\beta}^{\,\,\,0\rho\sigma} t'_{\rho\sigma} = 0 \label{eq:trho} \\
    \tilde{g}^{\nu i} \xi_i s_\nu + \tilde{g}^{\nu 0}s'_\nu = 0 \label{eq:snu}
\end{gather}
The first condition is the same as in \cite{Kovacs:2020ywu}, whilst the second condition is its Maxwell equivalent. 

These are the conditions we will need to show diagonalizability of $V^\pm$. Let us start with $V^+$. For contradiction, assume we have a non-trivial Jordan block so there exists $w = (u_I,u'_I)^T \in V^+$ such that 
\begin{equation} \label{eq:v}
    v = (M(\xi_i) - \xi_0)w \neq 0
\end{equation}
where $v = (T_I, \xi_0 T_I) \in V^+$ is an eigenvector of $M(\xi_i)$ with eigenvalue $\xi_0$. Write $T_I = (r_{\mu\nu}, p_\rho)$. Now, (\ref{eq:v}) is equivalent to the two equations
\begin{gather}
    u'_I = \xi_0 u_I + T_I \label{eq:uI} \\
    \mathcal{P}(\xi)^{IJ}u_J = - (2\xi_0 A + B)^{IJ}T_J \label{eq:Pfull}
\end{gather}
Note that we can substitute (\ref{eq:uI}) into the conditions (\ref{eq:trho}) and (\ref{eq:snu}) to get 
\begin{gather}
    \tilde{P}_{\beta}^{\,\,\,\mu\rho\sigma} \xi_\mu t_{\rho\sigma} + \tilde{P}_{\beta}^{\,\,\,0\rho\sigma} r_{\rho\sigma} = 0 \label{eq:trho2} \\
    \tilde{g}^{\nu \mu} \xi_\mu s_\nu + \tilde{g}^{\nu 0}p_\nu = 0 \label{eq:snu2}
\end{gather}
We can also rewrite (\ref{eq:Pfull}) by extracting the gauge-fixing terms:
\begin{equation}
    \mathcal{P}_{\star}(\xi)^{IJ}u_J = - \mathcal{P}_{GF}(\xi)^{IJ}u_J - (2\xi_0 A_{GF} + B_{GF})^{IJ}T_J - (2\xi_0 A_\star + B_\star)^{IJ}T_J \label{eq:gaug}
\end{equation}
The gauge-fixing terms are block-diagonal so they split into a gravitational part and a Maxwell part. In \cite{Kovacs:2020ywu}, it is shown that (\ref{eq:trho2}) implies the gravitational gauge-fixing part vanishes. The Maxwell parts vanish similarly since
\begin{align}
    - (\mathcal{P}_{GF}(\xi)^{\mu\nu}s_\nu &+ (2\xi_0 A_{GF} + B_{GF})^{\mu\nu}p_\nu) \notag\\
    &= \hat{g}^{\mu\gamma}\tilde{g}^{\nu\delta}\xi_\gamma \xi_\delta s_\nu +(2\xi_0\hat{g}^{\mu 0}\tilde{g}^{\nu 0} + \xi_i\hat{g}^{\mu i}\tilde{g}^{\nu 0} + \xi_i\hat{g}^{\mu 0}\tilde{g}^{\nu i})p_\nu \notag\\
    &= \hat{g}^{\mu\gamma}\tilde{g}^{\nu\delta}\xi_\gamma \xi_\delta s_\nu + \hat{g}^{\mu\gamma}\tilde{g}^{\nu 0}\xi_\gamma p_\nu + \hat{g}^{\mu 0}\tilde{g}^{\nu\gamma}\xi_\gamma p_\nu \notag\\
    &= \hat{g}^{\mu 0}\tilde{g}^{\nu\gamma}\xi_\gamma p_\nu \notag\\
    &= 0
\end{align}
where the third equality follows from (\ref{eq:snu2}) and the fourth from using the Case I condition on the eigenvector $v$. Hence all the gauge-fixing terms in (\ref{eq:gaug}) vanish, and contracting the remaining terms with $T_I^{\ast}$ gives
\begin{align}
    T_I^{\ast}\mathcal{P}_{\star}(\xi)^{IJ}u_J &= - T_I^{\ast}(2\xi_0 A_\star + B_\star)^{IJ}T_J \notag\\
    &= (v,v)_+
\end{align}
But by symmetry of $\mathcal{P}_{\star}(\xi)^{IJ}$, the left-hand side can be written as
\begin{equation}
    u_J\mathcal{P}_{\star}(\xi)^{JI}T_I^{\ast} = u_J(\mathcal{P}_{\star}(\xi)^{JI}T_I)^{\ast} = 0
\end{equation}
where the second equality follows because $v = (T_I, \xi_0 T_I)$ is a Case I eigenvector of $M$ and therefore $\mathcal{P}(\xi)^{IJ}T_J = 0$ and $\mathcal{P}_{GF}(\xi)^{IJ}T_J = 0$. But this implies that $(v,v)_+ = 0$, and since $(,)_+$ is positive definite this means $v=0$. This is a contradiction, and hence our assumption of a non-trivial Jordan block must be false, i.e., $M(\xi_i)$ must be diagonalizable in $V^+$. Repeating the arguments for $V^-$ gives us the same result. 

\subsection{Construction of Symmetrizer}

To summarize, we have found that $\tilde{V}^\pm$ and $\hat{V}^\pm$ are spaces of smoothly varying eigenvectors of $M(\xi_i)$ with real eigenvalues. We have also found that $V^\pm$ have bases of eigenvectors with real eigenvalues. However, the eigenvectors may not have smooth dependence on $\xi_i$ at points where eigenvalues cross, and so it is not obvious that the symmetrizer built from the eigenvectors will be smooth. 

Instead, as in \cite{Kovacs:2020ywu}, we will show that $H_\star^{\pm}$ is a symmetrizer for $M(\xi_i)$ within $V^{\pm}$. Let $v^{(1)} = (T_I^{(1)}, \xi^{(1)}_0 T^{(1)}_I)^T$ and $v^{(2)}= (T_I^{(2)}, \xi^{(2)}_0 T^{(2)}_I)^T$ be eigenvectors in $V^{\pm}$ with eigenvalues $\xi^{(1)}_0$ and $\xi^{(2)}_0$. Since all the eigenvalues associated with $V^{\pm}$ are real, we can take $v^{(1)}$ and $v^{(2)}$ to be real. Then
\begin{align}
    v^{(1)T}\left(M^T H_\star^{\pm} - H_\star^{\pm} M \right) v^{(2)} &= \left(\xi^{(1)}_0 - \xi^{(2)}_0\right)v^{(1)T}H_\star^{\pm}v^{(2)} \notag\\
    &= T_I^{(1)}\left( \left(\xi^{(1)2}_0 - \xi^{(2)2}_0\right) A_\star + \left(\xi^{(1)}_0 - \xi^{(2)}_0\right) B_\star\right)^{IJ}T^{(2)}_J \notag\\
    &= T_I^{(1)}\left(\mathcal{P}_{\star}(\xi^{(1)}) - \mathcal{P}_{\star}(\xi^{(2)}) \right)^{IJ}T^{(2)}_J = 0
\end{align}
The final equality follows because $v^{(1)}$ and $v^{(2)}$ are Case I eigenvectors, and so $\mathcal{P}_{\star}(\xi^{(1)})^{IJ}T_J^{(1)} = 0$ and $\mathcal{P}_{\star}(\xi^{(2)})^{IJ}T_J^{(2)} = 0$. Now, since eigenvectors form a basis of $V^{\pm}$, it follows that $H_\star^{\pm}$ is a symmetrizer for $M(\xi_i)$ within $V^{\pm}$. In particular, it depends smoothly on all its arguments and is positive definite.

We can construct the total symmetrizer in $V$ in an identical fashion to \cite{Kovacs:2020ywu}. Let $\{v^{\pm}_1, ..., v^{\pm}_4\}$ be a smooth basis for $V^{\pm}$, and let $\{\tilde{v}^{\pm}_1, ..., \tilde{v}^{\pm}_5\}$ and $\{\hat{v}^{\pm}_1, ..., \hat{v}^{\pm}_5\}$ be the smooth eigenvector bases constructed above for $\tilde{V}^{\pm}$ and $\hat{V}^{\pm}$. Let $S$ be the matrix whose columns are these basis vectors. Then the symmetriser is given by
\begin{equation}
    K(\xi_i) = (S^{-1})^T \begin{pmatrix}
        \mathcal{H}^+_\star & 0 & 0 & 0 & 0 & 0 \\
        0 & I_5 & 0 & 0 & 0 & 0 \\
        0 & 0 & I_5 & 0 & 0 & 0 \\
        0 & 0 & 0 & \mathcal{H}^-_\star & 0 & 0 \\
        0 & 0 & 0 & 0 & I_5 & 0 \\
        0 & 0 & 0 & 0 & 0 & I_5 
    \end{pmatrix}
    S^{-1}
\end{equation}
where $\mathcal{H}^\pm_\star$ are 4x4 matrices with components 
\begin{equation}
    (\mathcal{H}^\pm_\star)_{AB} = (v_A^\pm)^T H^\pm_\star v_B^\pm
\end{equation}
$K(\xi_i)$ has smooth dependence on $\xi_i$, the fields and all their derivatives, is positive definite and satisfies equation (\ref{eq:symmet}). Therefore, at weak coupling, our Einstein-Maxwell EFT is strongly hyperbolic and hence admits a well-posed initial value problem. 

\section{Conclusion} \label{Conclusion}

We have considered Einstein-Maxwell theory extended by the leading (4-derivative) effective field theory corrections. We have used the methods of \cite{Kovacs:2020ywu} to prove that the modified harmonic gauge formulation of this theory admits a well-posed initial value problem when the initial data is weakly coupled, i.e., when the 4-derivative terms in the equations of motion are initially small compared to the 2-derivative terms. Note that our result concerns {\it local} well-posedness, i.e., it guarantees existence of a solution only for a small interval of time. Over long time intervals, the fields may become large (e.g. if a singularity forms), in which case the theory would not be weakly coupled and well-posedness is likely to fail. From an EFT perspective this is fine because there is no reason to trust the theory if the fields become large. 

It is interesting to ask how large the higher derivative terms can become before strong hyperbolicity fails. Ref. \cite{Reall:2021} discussed this question for the case of the scalar-tensor EFT. In that case, it was shown that one can define a characteristic cone associated purely with the physical degrees of freedom. Using this cone one can define a notion of weak hyperbolicity that is independent of any gauge-fixing procedure. When this condition is satisfied, it was suggested that a sufficient condition for the modified harmonic gauge formulation to be strongly hyperbolic might be that the null cones of $\tilde{g}^{\mu\nu}$ and $\hat{g}^{\mu\nu}$ should lie strictly outside the characteristic cone. The same might be true for the Einstein-Maxwell EFT that we have considered. But determining whether or not this is the case, whether for scalar-tensor or Einstein-Maxwell EFT, will require new ideas. 

\section*{Acknowledgments}

We are grateful to \'Aron Kov\'acs for useful discussions. ID is supported by a STFC studentship. HSR is supported by STFC grant no. ST/T000694/1.

\section{Appendix}

\subsection{Principal Symbol}
Consider a theory of a pair $(g_{\mu\nu},A_\rho)$ defined by equations of motion $E^\mu = 0$, $E^{\mu\nu} = 0$. Let $\xi_\mu$ be an arbitrary covector. Then the principal symbol for these equations is a matrix
\begin{equation}
    \mathcal{P}(\xi)^{IJ} = \begin{pmatrix}
        \mathcal{P}_{gg}(\xi)^{\mu\nu\rho\sigma} & \mathcal{P}_{gm}(\xi)^{\mu\nu\rho} \\
        \mathcal{P}_{mg}(\xi)^{\mu\rho\sigma} & \mathcal{P}_{mm}(\xi)^{\mu\rho}
    \end{pmatrix}
\end{equation}
with elements defined by
\begin{align}
    \mathcal{P}_{gg}(\xi)^{\mu\nu\rho\sigma} = \frac{\partial E^{\mu\nu}}{\partial (\partial_\alpha\partial_\beta g_{\rho\sigma})}\xi_\alpha\xi_\beta \quad\quad&\quad\quad \mathcal{P}_{gm}(\xi)^{\mu\nu\rho} = \frac{\partial E^{\mu\nu}}{\partial (\partial_\alpha\partial_\beta A_{\rho})}\xi_\alpha\xi_\beta\\
    \mathcal{P}_{mg}(\xi)^{\mu\rho\sigma} = \frac{\partial E^{\mu}}{\partial (\partial_\alpha\partial_\beta g_{\rho\sigma})}\xi_\alpha\xi_\beta \quad\quad&\quad\quad \mathcal{P}_{mm}(\xi)^{\mu\rho} = \frac{\partial E^{\mu}}{\partial (\partial_\alpha\partial_\beta A_{\rho})}\xi_\alpha\xi_\beta
\end{align}
The matrix $\mathcal{P}(\xi)^{IJ}$ acts on the 14 dimensional vector space of "polarisation" vectors $(t_{\mu\nu},s_\rho)$ where $t_{\mu\nu}$ is symmetric, and so $I$ and $J$ run from 1 to 14 and refer to a basis of this vector space.

For our EFT in modified harmonic gauge with equations of motion $E^\mu_{mhg} = 0$, $E^{\mu\nu}_{mhg} = 0$, the principal symbol is given explicitly by
\begin{equation}
    \mathcal{P}(\xi)^{IJ} = \mathcal{P}^{EM}_{\star}(\xi)^{IJ} + \delta\mathcal{P}_{\star}(\xi)^{IJ} + \mathcal{P}_{GF}(\xi)^{IJ} 
\end{equation}
where 
\begin{align}
    \mathcal{P}_{\star}^{EM}(\xi)^{IJ} &= \begin{pmatrix}
        (-\frac{1}{2}g^{\alpha\beta}P^{\mu\nu\rho\sigma}+P_{\gamma}^{\,\,\,\alpha\mu\nu}P^{\gamma\beta\rho\sigma})\xi_\alpha \xi_\beta & 0 \\
        0 & (-g^{\mu\rho}g^{\alpha\beta}+g^{\mu\alpha}g^{\rho\beta})\xi_\alpha \xi_\beta
    \end{pmatrix}
\end{align}
\begin{align} \label{deltaP}
    \delta\mathcal{P}_{\star}(\xi)^{IJ} &= \begin{pmatrix}
        -c_3 T^{\mu\rho\lambda\alpha\nu\sigma\eta\beta}F_{\lambda\tau}F_\eta^{\,\,\,\tau}\xi_\alpha\xi_\beta & -2c_3 T^{\mu\gamma\lambda\alpha\nu\rho\eta\beta}\nabla_\eta F_{\lambda\gamma} \xi_\alpha\xi_\beta \\
        -2c_3 T^{\rho\gamma\lambda\alpha\sigma\mu\eta\beta}\nabla_\eta F_{\lambda\gamma} \xi_\alpha\xi_\beta & \left( 2c_3T^{\mu\alpha\lambda\eta\rho\beta\gamma\delta}R_{\lambda\eta\gamma\delta} + M^{\mu\rho\alpha\beta} \right)\xi_\alpha\xi_\beta
    \end{pmatrix}
\end{align}
\begin{equation}
    \mathcal{P}_{GF}(\xi)^{IJ} = \begin{pmatrix}
        -\hat{P}_{\gamma}^{\,\,\,\alpha\mu\nu}  \tilde{P}^{\gamma\beta\rho\sigma} \xi_\alpha\xi_\beta & 0\\
        0 & -\hat{g}^{\mu\alpha}\tilde{g}^{\rho\beta}\xi_\alpha \xi_\beta
    \end{pmatrix}
\end{equation}
where
\begin{align}
    P_{\alpha}^{\,\,\,\beta\mu\nu} &= \delta_\alpha^{(\mu}g^{\nu)\beta}-\frac{1}{2}\delta_\alpha^\beta g^{\mu\nu}\\
    \hat{P}_{\alpha}^{\,\,\,\beta\mu\nu} &= \delta_\alpha^{(\mu}\hat{g}^{\nu)\beta}-\frac{1}{2}\delta_\alpha^\beta \hat{g}^{\mu\nu}\\
    \tilde{P}_{\alpha}^{\,\,\,\beta\mu\nu} &= \delta_\alpha^{(\mu}\tilde{g}^{\nu)\beta}-\frac{1}{2}\delta_\alpha^\beta \tilde{g}^{\mu\nu}\\
    T^{\mu\rho\lambda\alpha\nu\sigma\eta\beta} &= \frac{1}{2}\left(\epsilon^{\mu\rho\lambda\alpha}\epsilon^{\nu\sigma\eta\beta} + \epsilon^{\nu\rho\lambda\alpha}\epsilon^{\mu\sigma\eta\beta}\right)
\end{align}
and
\begin{multline}
    M^{\mu\rho\alpha\beta} = 4\left(g^{\mu\rho}g^{\alpha\beta}-g^{\mu\alpha}g^{\rho\beta}\right)\partial_X f + 16F^{\mu\alpha}F^{\rho\beta}\partial_X^2 f\\ + 8\left( F^{\mu\alpha}\epsilon^{\rho\beta\gamma\delta}F_{\gamma\delta} + F^{\rho\beta}\epsilon^{\mu\alpha\gamma\delta}F_{\gamma\delta} \right)\partial_X\partial_Y f + 4\epsilon^{\mu\alpha\lambda\eta}\epsilon^{\rho\beta\gamma\delta}F_{\lambda\eta}F_{\gamma\delta}\partial^2_Y f
\end{multline}


\begin{thebibliography}{99}

\bibitem{Noakes:1983xd}
D.~R.~Noakes,
J. Math. Phys. \textbf{24}, 1846-1850 (1983)
doi:10.1063/1.525906

\bibitem{Flanagan:1996gw}
E.~E.~Flanagan and R.~M.~Wald,
Phys. Rev. D \textbf{54}, 6233-6283 (1996)
doi:10.1103/PhysRevD.54.6233
[arXiv:gr-qc/9602052 [gr-qc]].

\bibitem{Deser:1974cz}
S.~Deser and P.~van Nieuwenhuizen,
Phys. Rev. D \textbf{10}, 401 (1974)
doi:10.1103/PhysRevD.10.401

\bibitem{Jones:2019nev}
C.~R.~T.~Jones and B.~McPeak,
JHEP \textbf{06}, 140 (2020)
doi:10.1007/JHEP06(2020)140
[arXiv:1908.10452 [hep-th]].

\bibitem{Drummond:1979pp}
I.~T.~Drummond and S.~J.~Hathrell,
Phys. Rev. D \textbf{22}, 343 (1980)
doi:10.1103/PhysRevD.22.343

\bibitem{Horndeski:1980}
G.~Horndeski,
J. Math. Phys. \textbf{17} (1980)
doi:10.1063/1.522837

\bibitem{Abalos:2015gha}
F.~Abalos, F.~Carrasco, \'E.~Goulart and O.~Reula,
Phys. Rev. D \textbf{92}, no.8, 084024 (2015)
doi:10.1103/PhysRevD.92.084024
[arXiv:1507.02262 [gr-qc]].

\bibitem{Kidder:2001tz}
L.~E.~Kidder, M.~A.~Scheel and S.~A.~Teukolsky,
Phys. Rev. D \textbf{64}, 064017 (2001)
doi:10.1103/PhysRevD.64.064017
[arXiv:gr-qc/0105031 [gr-qc]].

\bibitem{Baumgarte:1998te}
T.~W.~Baumgarte and S.~L.~Shapiro,
Phys. Rev. D \textbf{59}, 024007 (1998)
doi:10.1103/PhysRevD.59.024007
[arXiv:gr-qc/9810065 [gr-qc]].

\bibitem{Shibata:1995we}
M.~Shibata and T.~Nakamura,
Phys. Rev. D \textbf{52}, 5428-5444 (1995)
doi:10.1103/PhysRevD.52.5428

\bibitem{Sarbach:2002bt}
O.~Sarbach, G.~Calabrese, J.~Pullin and M.~Tiglio,
Phys. Rev. D \textbf{66}, 064002 (2002)
doi:10.1103/PhysRevD.66.064002
[arXiv:gr-qc/0205064 [gr-qc]].

\bibitem{Nagy:2004td}
G.~Nagy, O.~E.~Ortiz and O.~A.~Reula,
Phys. Rev. D \textbf{70}, 044012 (2004)
doi:10.1103/PhysRevD.70.044012
[arXiv:gr-qc/0402123 [gr-qc]].

\bibitem{Papallo:2017qvl}
G.~Papallo and H.~S.~Reall,
Phys. Rev. D \textbf{96}, no.4, 044019 (2017)
doi:10.1103/PhysRevD.96.044019
[arXiv:1705.04370 [gr-qc]].

\bibitem{Papallo:2017ddx}
G.~Papallo,
Phys. Rev. D \textbf{96}, no.12, 124036 (2017)
doi:10.1103/PhysRevD.96.124036
[arXiv:1710.10155 [gr-qc]].

\bibitem{Kovacs:2019jqj}
\'A.~D.~Kov\'acs,
Phys. Rev. D \textbf{100}, no.2, 024005 (2019)
doi:10.1103/PhysRevD.100.024005
[arXiv:1904.00963 [gr-qc]].

\bibitem{Kovacs:2020pns}
\'A.~D.~Kov\'acs and H.~S.~Reall,
Phys. Rev. Lett. \textbf{124}, no.22, 221101 (2020)
doi:10.1103/PhysRevLett.124.221101
[arXiv:2003.04327 [gr-qc]].

\bibitem{Kovacs:2020ywu}
\'A.~D.~Kov\'acs and H.~S.~Reall,
Phys. Rev. D \textbf{101}, no.12, 124003 (2020)
doi:10.1103/PhysRevD.101.124003
[arXiv:2003.08398 [gr-qc]].

\bibitem{Taylor:1991}
M.~E.~Taylor,
 (Birkhäuser, Boston, MA, 1991).

\bibitem{Reall:2021}
H.~S.~Reall,
Phys. Rev. D \textbf{103}, no.8, 084027 (2021)
doi:10.1103/PhysRevD.103.084027

\bibitem{Bruhat:1952}
Y.~Fourès-Bruhat, 
Acta Math. \textbf{88}, 141 - 225, (1952) doi:10.1007/BF02392131

\end{thebibliography}
\end{document}